\documentclass[useAMS,usenatbib,usegraphicx,onecolumn]{mn2e}

\usepackage{latexsym}

\newcommand{\nc}{\newcommand}

\nc{\be}[1]{\begin{equation}\mbox{$\label{#1}$}}
\nc{\bea}[1]{\begin{eqnarray} \mbox{$\label{#1}$}}
\nc{\Section}[2]{\section{#2}\label{#1}}
\nc{\Bibitem}[1]{\bibitem{#1}}
\nc{\Label}[1]{\label{#1}}

\nc{\Mpc}{Mpc/h}

\nc{\eea}{\end{eqnarray}}
\nc{\ee}{\end{equation}}

\nc{\bdm}{\begin{displaymath}}
\nc{\edm}{\end{displaymath}}
\nc{\dpsty}{\displaystyle}
\nc{\bc}{\begin{center}}
\nc{\ec}{\end{center}}
\nc{\ba}{\begin{array}}
\nc{\ea}{\end{array}}
\nc{\bab}{\begin{abstract}}
\nc{\eab}{\end{abstract}}
\nc{\btab}{\begin{tabular}}
\nc{\etab}{\end{tabular}}
\nc{\bit}{\begin{itemize}}
\nc{\eit}{\end{itemize}}
\nc{\ben}{\begin{enumerate}}
\nc{\een}{\end{enumerate}}
\nc{\bfig}{\begin{figure}}
\nc{\efig}{\end{figure}}

\nc{\arreq}{&\!=\!&}
\nc{\arrmi}{&\!-\!&}
\nc{\arrpl}{&\!+\!&}
\nc{\arrap}{&\!\!\!\approx\!\!\!&}
\nc{\non}{\nonumber}
\nc{\align}{\!\!\!\!\!\!\!\!&&}

\def\lsim{\; \raise0.3ex\hbox{$<$\kern-0.75em
      \raise-1.1ex\hbox{$\sim$}}\; }
\def\gsim{\; \raise0.3ex\hbox{$>$\kern-0.75em
      \raise-1.1ex\hbox{$\sim$}}\; }

\nc{\DOT}{\hspace{-0.08in}{\bf .}\hspace{0.1in}}
\nc{\Laada}{\hbox {$\sqcap$ \kern -1em $\sqcup$}}
\nc\loota{{\scriptstyle\sqcap\kern-0.55em\hbox{$\scriptstyle\sqcup$}}}
\nc\Loota{{\sqcap\kern-0.65em\hbox{$\sqcup$}}}
\nc\laada{\Loota}
\nc{\qed}{\hskip 3em \hbox{\BOX} \vskip 2ex}

\nc{\real}{{\rm I \! R}}
\nc{\Z}{{\sf Z \!\!\! Z}}
\nc{\complex}{{\rm C\!\!\! {\sf I}\,\,}}
\def\bigid{\leavevmode\hbox{\small1\kern-3.8pt\normalsize1}}
\def\id{\leavevmode\hbox{\small1\kern-3.3pt\normalsize1}}
\nc{\slask}{\!\!\!/}
\nc{\bis}{{\prime\prime}}
\nc{\pa}{\partial}
\nc{\na}{\nabla}
\nc{\ra}{\rangle}
\nc{\goto}{\rightarrow}
\nc{\swap}{\leftrightarrow}

\nc{\EE}[1]{ \mbox{$\cdot10^{#1}$} }
\nc{\abs}[1]{\left|#1\right|}
\nc{\at}[2]{\left.#1\right|_{#2}}
\nc{\norm}[1]{\|#1\|}
\nc{\abscut}[2]{\Abs{#1}_{\scriptscriptstyle#2}}
\nc{\vek}[1]{{\rm\bf #1}}
\nc{\integral}[2]{\int\limits_{#1}^{#2}}
\nc{\inv}[1]{\frac{1}{#1}}
\nc{\dd}[2]{{{\partial #1}\over{\partial #2}}}
\nc{\ddd}[2]{{{{\partial}^2 #1}\over{\partial {#2}^2}}}
\nc{\dddd}[3]{{{{\partial}^2 #1}\over
    {\partial #2 \partial #3}}}
\nc{\dder}[2]{{{d #1}\over{d #2}}}
\nc{\ddder}[2]{{{d^2 #1}\over{d {#2}^2}}}
\nc{\dddder}[3]{{{d^2 #1}\over{d #2 d #3}}}
\nc{\dx}[1]{d\,^{#1}x}
\nc{\dy}[1]{d\,^{#1}y}
\nc{\dz}[1]{d\,^{#1}z}
\nc{\dl}[1]{\frac{d\,^{#1}l}{(2\pi)^{#1}}}
\nc{\dk}[1]{\frac{d\,^{#1}k}{(2\pi)^{#1}}}
\nc{\dq}[1]{\frac{d\,^{#1}q}{(2\pi)^{#1}}}

\nc{\bfT}{{\bf T }}

\nc{\cA}{{\cal A}}
\nc{\cB}{{\cal B}}
\nc{\cD}{{\cal D}}
\nc{\cE}{{\cal E}}
\nc{\cG}{{\cal G}}
\nc{\cH}{{\cal H}}
\nc{\cL}{{\cal L}}
\nc{\cO}{{\cal O}}
\nc{\cT}{{\cal T}}
\nc{\cN}{{\cal N}}
\nc{\cR}{{\cal R}}
\nc{\rvac}[1]{|{\cal O}#1\rangle}
\nc{\lvac}[1]{\langle{\cal O}#1|}
\nc{\rvacb}[1]{|{\cal O}_\beta #1\rangle}
\nc{\lvacb}[1]{\langle{\cal O}_\beta #1 |}
\nc{\bb}{\bar{\beta}}
\nc{\bt}{\tilde{\beta}}
\nc{\ctH}{\tilde{\cal H}}
\nc{\chH}{\hat{\cal H}}
\nc{\1}{\aa}
\nc{\2}{\"{a}}
\nc{\3}{\"{o}}
\nc{\4}{\AA}
\nc{\5}{\"{A}}
\nc{\6}{\"{O}}
\nc{\al}{\alpha}
\nc{\g}{\gamma}
\nc{\Del}{\Delta}
\nc{\e}{\textrm{e}}
\nc{\eps}{\epsilon}
\nc{\lam}{\lambda}
\nc{\Om}{\Omega}
\nc{\ve}{\varepsilon}
\nc{\mn}{{\mu\nu}}
\nc{\vp}{\varphi}

\nc{\apj}[3]{{#2}, ApJ, {#1}, {#3}}
\nc{\astro}[3]{{#2}, AJ, {#1}, {#3}}
\nc{\mnras}[3]{{#2}, MNRAS, {#1}, {#3}}
\nc{\ncim}[3]{{  Nuov.\ Cim.\ }{{ #1} {(#2)} {#3}}}
\nc{\np}[3]{{  Nucl.\ Phys.\ }{{ #1} {(#2)} {#3}}}
\nc{\pr}[3]{{#2}, Phys. Rev., {#1}, {#3}}
\nc{\pl}[3]{{#2}, Phys. Lett., {#1}, {#3}}
\nc{\prep}[3]{{#2}, Phys. Rep., {#1}, {#3}}
\nc{\rmp}[3]{{#2}, Rev. Mod. Phys., {#1}, {#3}}

\nc{\rf}[1]{(\ref{#1})}
\nc{\nn}{\nonumber \\*}
\nc{\bfB}{\bf{B}}
\nc{\bfv}{\bf{v}}
\nc{\bfx}{\bf{x}}
\nc{\bfy}{\bf{y}}
\nc{\vx}{\vec{x}}
\nc{\vy}{\vec{y}}
\nc{\oB}{\overline{B}}
\nc{\oI}{\overline{I}}
\nc{\oR}{\overline{R}}
\nc{\rar}{\rightarrow}
\nc{\ti}{\times}
\nc{\slsh}{\hskip-5pt/}
\nc{\sm}{Standard~Model~}
\nc{\MP}{M_{\rm Pl}}
\nc{\tp}{t_{\rm Pl}}

\nc{\pmin}{p_{\rm min}}
\nc{\pmax}{p_{\rm max}}
\nc{\fo}{f_0}
\nc{\foi}{f_{0,i}\,}
\nc{\fop}{f_0^P}
\nc{\fou}{f_0^U}

\nc{\eff}{{\rm eff}}
\nc{\MT}{M_{\rm T}}
\nc{\ML}{M_{\rm L}}
\nc{\kk}{\vek{k}}
\nc{\pp}{{\rm p}}
\nc{\pt}{\partial_t}
\nc{\half}{{1\over 2}}
\nc{\w}{\omega}
\nc{\uhat}{\hat{U}_\w}

\nc{\etal}{\mbox{et al.}}
\nc{\ie}{{i.e. }}
\nc{\eg}{{e.g. }}
\nc{\trh}{T_{\rm RH}}
\nc{\ad}{{a'\over a}}
\nc{\bd}{{b'\over b}}
\nc{\Rd}{{R'\over R}}
\nc{\diag}{{\textrm{diag}}}
\nc{\mato}[1]{\tilde{#1}}
\nc{\sech}{\textrm{sech}}
\nc{\I}{\textrm{I}}
\nc{\II}{\textrm{II}}
\nc{\III}{\textrm{III}}
\nc{\vev}[1]{\langle #1 \rangle}
\nc{\hyper}{\,\; F_{1{\hskip -16pt}2}{\hskip 11pt}}
\nc{\brhom}{\overline{\rho}_M}
\nc{\rhob}{\overline{\rho}}
\nc{\Pb}{\overline{P}}
\nc{\bH}{\overline{H}}

\title[]{Large scale structure in non-standard cosmologies}

\author[ T. Multam\"aki, E. Gazta\~naga and  M. Manera]
{Tuomas Multam\"aki$^1$\thanks{tuomas@ecm.ub.es}
Enrique Gazta\~naga,$^2$\thanks{gazta@ieec.fcr.es}
Marc Manera,$^2$\thanks{manera@ieec.fcr.es}
\\
$^1$ Departament E.C.M. and C.E.R. en Astrofisica, Fisica de Particules 
i Cosmologia,\\ Universitat de Barcelona, Diagonal
647, 08028 Barcelona, Spain\\
$^2$ Insitut d'Estudis Espacials de Catalunya, IEEC/CSIC, Gran
Capit\'an 2-4, 08034 Barcelona, Spain}

\date{Draft version 21 March 2003}

\pagerange{\pageref{firstpage}--\pageref{lastpage}} \pubyear{2002}

\begin{document}

\label{firstpage}

\maketitle

\begin{abstract}

We study the growth of large scale structure in two
recently proposed non-standard cosmological models: the brane induced 
gravity model of Dvali, Gabadadze and Porrati (DGP) and the Cardassian
models of Freese and Lewis. A general formalism for
calculating the growth of fluctuations in models with a non-standard
Friedman equation and a normal continuity equation of energy density is
discussed.
Both linear and non-linear growth are studied, together with their
observational signatures on higher order statistics and abundance of
collapsed objects. In general, models which show similar cosmic acceleration 
at $z\simeq 1$,  can produce quite different normalization for large
scale density fluctuations,\ie$\sigma_8$, cluster abundance or
higher order statistics, such as the normalized skewness $S_3$, which is
independent of the linear normalization. 
For example, for a flat universe with $\Omega_m \simeq 0.22$,
DGP and standard Cardassian cosmologies predict about 2 and 3 times more 
clusters respectively
than the standard $\Lambda$  model at $z=1.5$.  When normalized to CMB 
fluctuations the $\sigma_8$  amplitude turns out to be lower 
by a few tens of a percent. We also find that, for a limited
red-shift range, the linear growth rate can be faster 
in some models (eg modified polytropic Cardassian with $q>1$)
than in the Einstein-deSitter universe. The value of the skewness 
$S_3$  is found to have up to  $\simeq 10\%$ percent variations
(up or down) from model to model.

\end{abstract}

\begin{keywords}
cosmology: theory -- cosmological parameters -- large-scale structure of the Universe
\end{keywords}

\section{Introduction}
The suggestion that the universe is undergoing late time acceleration
(eg Efstathiou, Sutherland \& Maddox 1990;
Riess et al. 1998; Perlmutter et al. 1999)
has provoked a number of cosmological models that can reproduce the
appropriate evolution. The obvious 
solution is clearly to add the cosmological constant
to the equations determining the evolution of the universe
(see \eg Weinberg 1989). This,
as is very well known, is problematic from the point of view of particle
physics due to the smallness of the required constant. It is then
interesting to consider scenarios where instead of the cosmological
constant, there is a new type of mechanism that can explain the
acceleration of the universe. In particular, in this paper we
consider two recent suggestions: the Dvali-Gabadadze-Poratti (DGP) brane
induced gravity -model (Dvali, Gabadadze \& Porrati 2000) and the 
Cardassian model(s) of Freese and Lewis (Freese \& Lewis 2002).
Both models lead to late-time acceleration with a non-standard
Friedmann equation and with no explicit cosmological constant.
The two scenarios are, however, fundamentally different in their
properties, due to the fact that the DGP-scenario is a truly higher 
dimensional scenario while the Cardassian is an effective
description.

Both of these scenarios have been studied from the point of
view of observational constraints (Deffayet et al. 2002; Wang 2003;
Gondolo \& Freese 2002; Sen \& Sen 2002, 2003; Zhu \& Fujimoto 2002, 2003)
mainly coming from the SNIa (Riess et al. 1998; Perlmutter et al. 1999) 
and CMB data. The constrains are not strong enough to exclude either of the
models as an alternative to the standard $\Lambda$-cosmology.
Are they compatible with other current observations, such as the
large scale structure?

In this paper we study the DGP and Cardassian scenarios from 
the point of view
of large scale structure growth. Since these scenarios
modify gravity on large scales, the growth is non-standard. 
The linear growth of perturbations in the original Cardassian
scenario was also briefly considered in (Gondolo \& Freese 2002).
Our approach is much like that in (Gazta\~naga \& Lobo 2001), where
the gravitational growth was studied in a number
of non-standard scenarios. However, the Einstein's equations
are modified on large scales, so we must alter our
approach accordingly.

This paper is organized as follows: In Section 2 we introduce
the two scenarios and briefly consider the experimental
constraints. In Section 3 we introduce the formalism of
studying gravitational collapse in a matter dominated 
cosmology where energy density of matter is conserved
but the Friedmann equation is arbitrary. The linear
and non-linear growth of perturbations in a standard
$\Lambda$-cosmology is also discussed in this Section.
In Sections 4 and 5 we study gravitational collapse
in the DGP and Cardassian scenarios and present the
results. Observational constraints coming from the linear
and non-linear aspects of structure formation are briefly
considered in Section 6. The paper is concluded in Section 7.
In the Appendix, the growth of perturbations in the 
original Cardassian model is studied analytically.

\section{The two scenarios}

\subsection{Brane induced gravity model}

The brane induced gravity model (Dvali et al. 2000) offers an
alternative explanation to the observed acceleration
by a large scale modification of gravity due to the presence
of an extra dimension (Deffayet et al. 2002a,b).
In this scenario there
is hence no need for an explicit non-zero cosmological constant.
At large enough scales gravity sees the full space-time, 
\ie the brane {\it and} the bulk,
and is therefore modified from the standard $1/r^2$ law\footnote{
It was later realized that in the DGP scenario gravity {\it is}
modified also on small scales and may be detected in anomalous
precession of orbiting bodies in the Solar System (Lue \& Starkman 2002; 
Dvali, Gruzinov \& Zaldarriaga 2002)}.

In the DGP-scenario the Friedmann equation on the brane is (Deffayet et al. 
2002a):
\be{fried} 
H^2=\Big(\sqrt{\kappa^2\rho+{1\over R^2}}+{1\over R}\Big)^2+{K\over a^2}, 
\ee where $H\equiv\dot{a}/a$,
$K=0,\pm 1$ is the curvature constant, $a$ the scale
factor of the universe and we have defined
$\kappa^2=(8\pi G)/3$, $R\equiv 2 r_c$.
The evolution of the scale factor is standard as long as the energy
density of matter dominates \ie $\kappa^2\rho\gg 1/(4 r_c^2)$.
Since $\rho$ decreases with time due to the expansion of the
universe, the $r_c^{-2}$-term becomes dominant at some point, and
acting as an effective cosmological constant, it leads
to an accelerating universe. At late times the scale factor grows
exponentially, $a\sim\exp(t/r_c)$. 
The continuity equation is unchanged, 
\be{cont}
\dot{\rho}+3 H (\rho+p)=0.
\ee

We are interested in the growth of large scale structure which 
takes place in a matter dominated universe and hence we 
assume that $\rho\gg p$ and therefore the continuity equation tells us 
that $\rho\sim a^{-3}$. We can view the DGP-scenario as standard
cosmology with a perfect fluid with the normal properties. The
only difference is the non-standard Friedmann equation.

We define the cosmological quantities as usual
\bea{omdefs}
\Omega_M & \equiv & {\kappa^2\rho_0\over H_0^2}\\
\Omega_R & \equiv & {1\over R^2H_0^2}\\
 \Omega_K & \equiv & {K\over H_0^2}, \eea
where $H_0$ is the current Hubble rate and $\rho_0$ the current
density of matter. Since we are interested in cosmology in the
matter dominated universe, we can relate $\Omega_M$ to $\rho$ by
(we take $a_0\equiv 1$)
\be{omegamrho}
\Omega_M={\kappa^2\rho_0\over H_0^2}={\rho a^3\over H_0^2}.
\ee

The Friedmann equation, (\ref{fried}), can then be written as
\be{fried2}
H^2=H_0^2\Big((\sqrt{{\Omega_M\over a^3}+\Omega_R}+\sqrt{\Omega_R})^2+
{\Omega_K\over a^2}). 
\ee
Note that the normalization condition differs from the usual one,
\be{norm}
1=\Big(\sqrt{\Omega_M+\Omega_R}+\sqrt{\Omega_R}\Big)^2+\Omega_K.
\ee
From now on we assume that we live in a flat universe, $K=0$, so that
Friedmann equation can be written as:
\be{fried3}
H^2 = H_0^2\Big({\Omega_M\over a^3}
+2 \Omega_R(1+\sqrt{1+{\Omega_M\over\Omega_R a^3}})\Big).
\ee
The normalization condition simplifies to
\be{flatnorm}
\Omega_M+2\sqrt{\Omega_R}=1,
\ee
from where it is clear that in order to have $\Omega_M>0$, $\Omega_R$ 
must be restricted to the range
\be{flatomr}
0\leq \Omega_R < \frac 14.
\ee

In (Deffayet et al. 2002a) this model was tested
with SNIa and the CMB data. It was found that the model is
in agreement with data\footnote{This result was, however, disputed 
by Avelino \& Martins (2002) who argued that this scenario is already
strongly disfavored.}
 and the preferred parameter values
for a flat universe are
\be{dgpparam}
\Omega_M=0.18^{+0.07}_{-0.06},\ \Omega_R=0.17^{+0.03}_{-0.02}.
\ee

\subsection{Cardassian models}
In the Cardassian\footnote{Humanoid-like race from Star Trek, see
e.g. www.startrek.com}
models (Freese \& Lewis 2002; Gondolo \& Freese 2002) the Friedmann equation has the general form
\be{carfried}
H^2=g(\rho_M),
\ee
where the $\rho_M$ is the energy density of ordinary matter and
radiation. The universe is assumed to be flat and that there is
no new type of matter nor a non-zero cosmological constant. 
The function $g$ is assumed to approach
the standard form, $\kappa^2\rho$ at early times, including nucleosynthesis,
and at late times, $z<\cO(1)$, to give accelerated expansion
in accordance with the supernova observations \cite{sna}.
Since the behavior of the function is different at different
values of $\rho_M$, there is an associated scale, $\rho_{C}$, or
red-shift, $z_{eq}$, in the function $g$ that determines when
the evolution is standard and when the non-standard terms begin to
dominate. The exact form of the model can hence vary, as long
as it satisfies the aforementioned constraints. Some alternative
forms of the Friedman equation are presented in (Freese 2002).

As an example, consider the following (original Cardassian) form 
of $g(\rho)$ (Freese \& Lewis 2002) (we omit the subscript $M$ from now on):
\be{car1}
H^2=\kappa^2\rho+B\rho^n,\ n<\frac 23.
\ee
At early times the universe is dominated
by the $\kappa^2\rho$-term (provided that $B$ is small enough at
the time of interest) and at late times the $\rho^n$-term
becomes significant, providing acceleration compared to the standard
case.  The term of the form $\rho^n$ in the Friedmann 
equation, and hence the Cardassian model(s), is motivated by 
considering our universe as a brane embedded in extra dimensions
(Chung \& Freese 2000) (this idea was, however, critically reviewed
in (Cline \& Vinet 2002)).
In terms of the scale $\rho_C$, Eq. (\ref{car1}) can be written
as
\be{car2}
H^2=\kappa^2\rho[1+({\rho\over\rho_C})^{n-1}],
\ee
hence $B=\kappa^2\rho_C^{1-n}$. In a matter dominated
universe this is conveniently parametrized by the red-shift 
at which the two terms are equal, $z_{eq}$, 
\be{car3}
H^2=\kappa^2\rho[1+(1+z_{eq})^{3(1-n)}({\rho\over\rho_0})^{n-1}],
\ee
where $\rho_0$ is the current energy density of matter.

The requirement that $n<\frac 23$ is actually more universal as one
can see from considering the acceleration of the scale factor from
the general Cardassian form, Eq. (\ref{carfried}):
\be{caracc}
{\ddot{a}\over a}=g(\rho)-{3\over 2}\rho g'(\rho),
\ee
where it has been assumed that the energy density of ordinary matter
is conserved and has no pressure \ie $\dot{\rho}+3H\rho=0$.
If we wish to have late time acceleration, $\ddot{a}/a$ must be
greater than zero at late times (when the assumption that $P=0$
becomes more and more exact) and hence at late times the inequality
\be{caracclate}
g(\rho)_{late}<B\rho^{\frac 23}
\ee
must hold. Therefore, in order to have late time acceleration, the
Cardassian function, $g(\rho)$, must grow more slowly than 
$\rho^{\frac 23}$ at late times.

Recently, another Cardassian model has been studied in
more detail  with respect to the future SNIa
observations (Wang et al. 2003). In the Modified Polytropic 
Cardassian (MPC) model, the Friedman equation is
\be{mpcequ}
H^2=\kappa^2\rho[1+\Big({\rho\over\rho_C}\Big)^{q(n-1)}]^{1/q},
\ee
where $\rho_C$ is again the energy density of matter at which the
non-standard terms begin to dominate and $q>0$ is a parameter\footnote{Incidentally, the Q-continuum is another race from Star
Trek, existing in extra dimensions (www.startrek.com)}. 
The original Cardassian model is a special case
of the MPC model with q=1 and hence we shall concentrate on this more
general model in this paper.
The Friedmann equation in the MPC-scenario in a matter dominated
universe can equally well be written 
in terms of the red-shift at which the Cardassian terms start to dominate:
\be{mpcequ2}
H^2=\kappa^2\,\rho{\Big( 1 + 
         {( 1 + {z_{eq}} ) }^
           {3\,( 1 - n ) \,q}\,
          {( \frac{\rho}{{\rho_0}} ) }^
           {(n-1)q} \Big) }^{\frac{1}{q}}=H_0^2{\Omega_M\over a^3}
\Big(1+(1+z_{eq})^{3(1-n)q}a^{3(1-n)q}\Big)^{1\over q},
\ee
where $\rho_0$ is the energy density of matter today.

In a flat universe, which we assume to the be case throughout this paper,
the observed matter density of the universe, $\Omega_M^{obs}$, 
can be related to $z_{eq}$ by
(Wang et al. 2003) by
\be{mpcobs}
1+z_{eq}=[(\Omega_M^{obs})^{-q}-1]^{1/(3q(1-n))}.
\ee
The MPC-model is constrained by the supernova observations as well
as the CMB (Wang et al. 2003).
The experimentally allowed $(n,q)$ parameter 
space is large and there is a degeneracy along the $q$ axis at
for $q\gsim 10$ (when $\Omega_M^{obs}=0.3$) (Wang et al. 2003).

The original Cardassian model has also been constrained
in other works: in (Zhu \& Fujimoto 2002) the angular size of compact radio sources at different red-shifts was considered and in (Sen \& Sen 2002)
by the SNIa, as well as in (Zhu \& Fujimoto 2003), and CMB data. 


\section{Gravitational growth}

Before studying the gravitational growth of structures in the two
aforementioned scenarios, we first briefly introduce the
usual formalism.

Raychaudhuri's equation for a shear free fluid with four-velocity
$u^{\mu}$, is
\be{ray}
\dot{\Theta}+{1\over 3}\Theta^2=R_{\mu\nu}u^{\mu}u^{\nu}
\ee
where 
\be{thet1}
\Theta \equiv \nabla_{\mu}u^{\mu}.
\ee
Choosing the coordinate system such that
the four-velocity of the fluid, $u^{\mu}$, is
\be{velo}
u^{\mu}=(1,\dot{a}{\bf x}+{\bf v}),
\ee
where ${\bf v}$ is the peculiar velocity.
Hence,
\be{thet2}
\Theta=3{\dot{a}\over a}+{\theta\over a},
\ee
where we have defined $\theta\equiv\nabla\cdot{\bf v}$.
Assuming that the geometry of the universe is of the standard FRW-form,
the Raychaudhuri's equation simplifies to
\be{ray2}
\dot{\Theta}+{1\over 3}\Theta^2=3(\dot{H}+H^2).
\ee
Note, that in order to study the growth of perturbations, 
on the LHS of Eq. (\ref{ray2}), we are using the background
quantities and on the RHS the local, perturbed quantities. 
Hence, using (\ref{thet2}), we get
\be{ray3}
{\dot{\theta}\over a}+{\theta\over a}\overline{H}+\frac 13{\theta^2\over
a^2}=3(\dot{H}+H^2)-3(\dot{\bH}+\bH^2).
\ee
Up to this point, we have not assumed anything but that
we live in a FRW universe filled with a perfect fluid and hence
Eq. (\ref{ray3}) applies both to the DGP- and Cardassian-scenarios. 
In particular, we have not assumed any connection between the geometry 
and the energy content but simply rewritten the Raychaudhuri's equation in
terms of the scale factor. This is useful since now we can study
the evolution of the density perturbations armed with the Friedmann
equation and the continuity equation, without having to concern us
with the Einstein's equations. Obviously, these considerations
make only sense on large scales where the effects of the non-standard
cosmology can be seen.

In order to have a deeper understanding of the different scenarios,
we recall the growth of perturbations in the standard scenario
with a non-zero cosmological constant.


\subsection{Standard scenario with $\Lambda\neq 0$}\label{lamuni}

In the standard scenario, we know the Einstein's equations so it
is instructive to see how one arrives to the same result by using
them and Eq. (\ref{ray3}) directly. 

From the Einstein's equations
\be{ees}
R_{\mu\nu}-{1\over 2}g_{\mu\nu}\cR=-\kappa^2T_{\mu\nu}+\Lambda g_{\mu\nu}
\ee
and the energy-momentum tensor of an ideal fluid
\be{tmunu}
T_{\mu\nu}=Pg_{\mu\nu}+(P+\rho)u_{\mu}u_{\nu},
\ee
it can be verified that in the FRW-metric
\be{rayrhs}
R_{\mu\nu}u^{\mu}u^{\nu}=-\frac 32\kappa^2(\rho+3P)+\Lambda.
\ee
On the other hand, by using the Friedmann equation
\be{stanfried}
H^2=\kappa^2\rho+{\Lambda\over 3}
\ee
and the continuity equation 
\be{stancont}
\dot{\rho}+3H(\rho+P)=0,
\ee
we see that
\be{standhs}
3(\dot{H}+H^2)=-\frac 32\kappa^2(\rho+3P)+\Lambda
\ee
\ie we arrive to the same equation using only the Friedmann and
the continuity equations.
The Raychaudhuri's equation in the standard case
hence takes the form
\be{stanray3}
{\dot{\theta}\over a}+{\theta\over a}\overline{H}+\frac 13{\theta^2\over
a^2}=-\frac 32\kappa^2(\rho-\rhob+3(P-\Pb)).
\ee

In a matter dominated universe, the continuity equation 
for a non-relativistic ($\rho\gg p$) fluid can 
be written as (Peebles 1993) 
\be{fluid}
{d\delta\over d\tau}+(1+\delta)\theta=0,
\ee
where 
\be{delta}
\delta(\tau,{\bf x})={\rho(\tau,{\bf x})\over \bar{\rho}(\tau)}-1
\ee
is the local density contrast and we have switched to conformal time
$dt=ad\tau$.
Using (\ref{fluid}), (\ref{delta}) and (\ref{stanray3}) we see that
in terms of the conformal time we get:
\be{standenspe}
{d^2\delta\over d\tau^2}+\cH{d\delta\over d\tau}-\frac 43{1\over
1+\delta}({d\delta\over d\tau})^2=\frac 32\kappa^2(1+\delta)\delta\rhob a^2,
\ee
where $\cH\equiv d(\ln\, a)/d\tau$.
Rescaling the time variable once more, $\eta=\ln(a)$, we arrive at
\be{standenspe2}
{d^2\delta\over d\eta^2}+(2+{\dot{\bH}\over \bH^2}){d\delta\over d\eta}
-\frac 43{1\over 1+\delta}({d\delta\over d\tau})^2=\frac 32\kappa^2
(1+\delta)\delta{\rhob\over\bH^2}.
\ee

Using the notations (\ref{omdefs}), and recalling
that in the matter dominated regime it follows from the continuity
equation, Eq. (\ref{stancont}), that $\rho\sim a^{-3}$, the
Friedmann equation can be written as
\be{stanfried2}
H^2=H_0^2[{\Omega_M\over a^3}+\Omega_{\Lambda}].
\ee
Using this in Eq. (\ref{standenspe2}) gives us the well known result
\be{standenspe3}
{d^2\delta\over d\eta^2}+\Big[2-\frac
32{\Omega_M\over\Omega_M+a^3\Omega_{\Lambda}}\Big]{d\delta\over
d\eta}-\frac 43{1\over 1+\delta}\Big({d\delta\over d\tau}\Big)^2=
\frac 32(1+\delta)\delta{\Omega_M\over\Omega_M+a^3\Omega_{\Lambda}}.
\ee

In order to determine how a small perturbation grows with time at
different orders in perturbation theory, we
expand $\delta$ as:
\be{deltaexp}
\delta=\sum_{i=1}^{\infty}\delta_i=\sum_{i=1}^{\infty}{D_i(\eta)\over i!}\delta_0^i,
\ee
where $\delta_0$ is the small perturbation (and the expansion parameter).
Using the expansion, we get the linear equation
\be{lin}
D_1''+(2-\frac32{\Omega_M\over\Omega_M+a^3\Omega_{\Lambda}})D_1'-
\frac32{\Omega_M\over\Omega_M+a^3\Omega_{\Lambda}}D_1=0,
\ee
which in an Einstein-deSitter (EdS) universe ($\Omega_M=1,\
\Omega_{\Lambda}=0$) has the well known solution
\be{linsol}
D_1(\eta)=c_1e^{\eta}+c_2e^{-3\eta/2}.
\ee
The solution to the linear equation in the general case with non-zero 
$\Lambda$, can be 
expressed in terms of the hyper-geometric function as
\be{lamunisol}
D_1=c_1{\sqrt{1+{\Omega_{\Lambda}\over\Omega_M}a^3}\over a^{3/2}}+
c_2\ \hyper \Big[1,{1\over 3},{11\over 6},- {\Omega_{\Lambda}\over\Omega_M}a^3\Big]a.
\ee

The second order equation is
\be{2ndord}
D_2''+(2-\frac32{\Omega_M\over\Omega_M+a^3\Omega_{\Lambda}})D_2'-
\frac32{\Omega_M\over\Omega_M+a^3\Omega_{\Lambda}}D_2-
\frac 83(D_1')^2-3D_1^2{\Omega_M\over\Omega_M+a^3\Omega_{\Lambda}}=0,
\ee
where it is understood that the linear solution is
substituted. Similarly one can recursively go on to arbitrary order.

The second order equation determines how Gaussian initial 
conditions develop non-Gaussian features and can be related to the
skewness of the density field at large scale. The $q$-order moments
of the fluctuating field are related to the perturbations by
\cite{bcgs}
\be{jmoment}
m_q\equiv \vev{\delta^q},
\ee
which in term can be related to the connected moments, or cumulants, 
$\bar{\xi}_q$. The normalized skewness is given by
\cite{bcgs}
\be{skew}
S_3={\bar{\xi}_3\over \bar{\xi}_2^2}={m_3\over m_2^2},
\ee
which can be written in terms of the first and second order
perturbations. For example at leading order:
\be{m3}
 m_3 = \vev{\delta^3} \simeq  \vev{\delta_1^3}
+3\vev{\delta_2~\delta_1^2} + \dots
\ee
For Gaussian perturbations $\vev{\delta_1^3}=0$, so that we get:
\be{pertskew}
S_3=3{D_2\over D_1^2}.
\ee
In an Einstein-deSitter universe this coefficient can be calculated
exactly and it is $S_3^{EdS}=34/7\approx 4.86$.

\begin{figure}
\includegraphics[width=70mm,angle=-90]{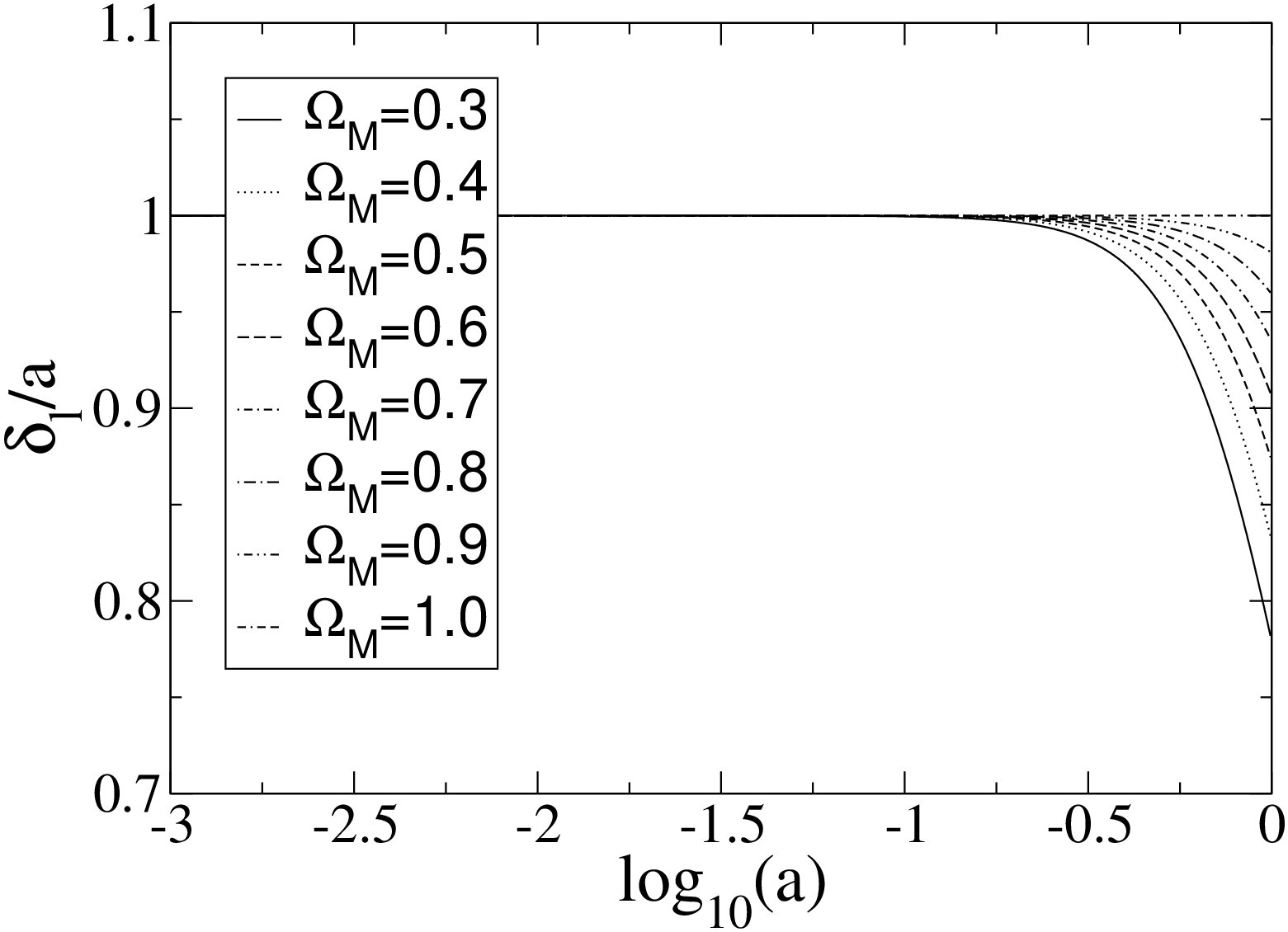}
\includegraphics[width=70mm,angle=-90]{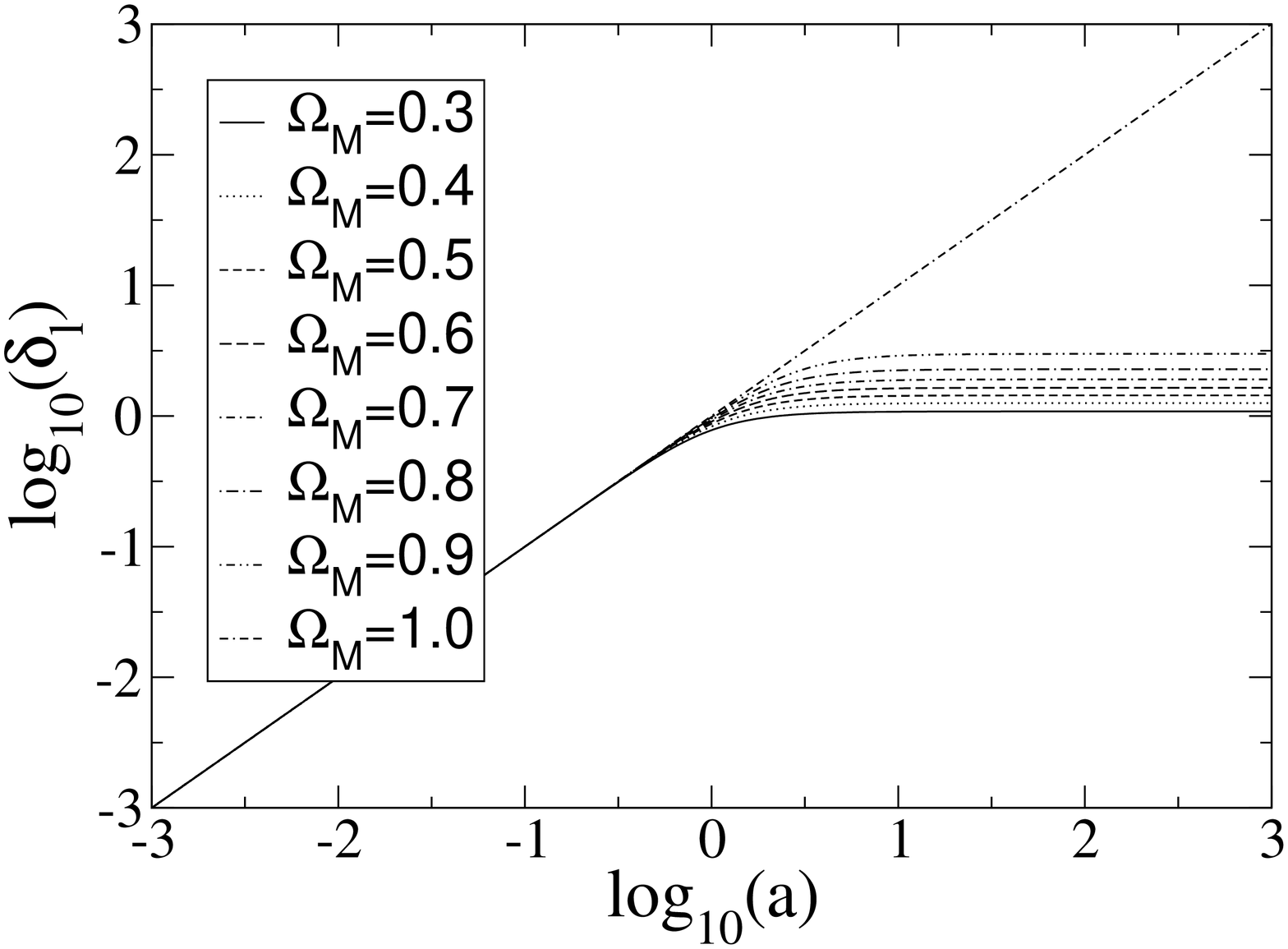}
\caption{Linear growth with for different values of $\Omega_M$}
\label{stfig}
\end{figure}

\begin{figure}
\includegraphics[width=70mm,angle=-90]{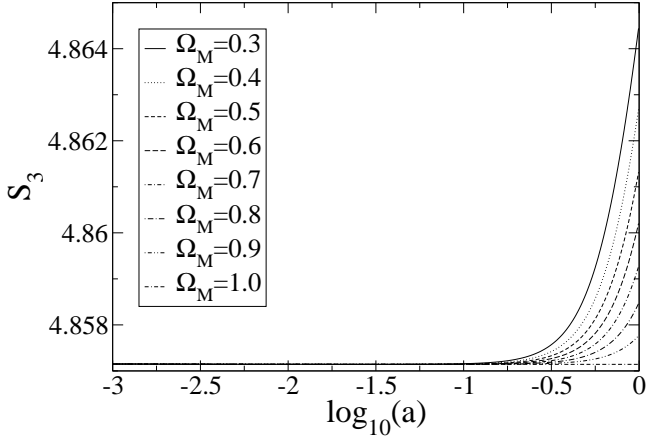}
\caption{Non-linear growth for different values of $\Omega_M$}
\label{stfig2}
\end{figure}

To illustrate the effect of the cosmological constant, we have solved
the linear and second order equations numerically. The linear
evolution and the evolution of $S_3$ are shown in Figs \ref{stfig} and
\ref{stfig2}.

We see that the effect on the linear growth factor is significant as
perturbations grow less when the cosmological constant is large
compared to the energy density of matter. The effect of the
cosmological constant on linear growth is well represented 
by the other figure where future evolution is shown. With a non-zero
$\Omega_\Lambda$, $\delta_l$ freezes and structures stop growing
in the future.

The effect on $S_3$ is very small, as expected, so that even 
when $\Omega_M=0.3$, $\Delta S_3< 1\%$. Such a small change
is clearly out of reach of present day observations as is
discussed in more detail in Section 6.


\subsection{Gravitational growth with a general Friedmann equation}
In a general case with the Raychaudhuri's equation given by
Eq. (\ref{ray3}), the equation governing large scale structure growth in the
matter dominated universe can be
derived following similar steps like in the standard case. The result
is
\be{gendenspe}
{d^2\delta\over d\eta^2}+(2+{\dot{\bH}\over \bH^2}){d\delta\over d\eta}
-\frac 43{1\over 1+\delta}({d\delta\over
d\tau})^2=-3{1+\delta\over\bH^2}\Big((\dot{H}+H^2)-(\dot{\bH}+\bH^2)\Big),
\ee
where again $\eta=\ln(a)$.
Obviously one cannot progress further unless the continuity equation
is known so that $\dot{H}$ can be calculated. With the continuity
equation one can then express $\rho$ in terms of $\bar{\rho}$ and
$\delta$ using Eq. (\ref{delta}). The resulting equation
can be expressed in terms of cosmological quantities, 
$\Omega_M,\ \Omega_{\Lambda},$..., by writing the Friedmann equation
in terms of $\Omega$s and noting that from Eq. (\ref{omegamrho}) we
get
\be{generhobar}
\bar{\rho}={H_0^2\over\kappa^2 a^3}\Omega_M.
\ee
We then have an equation determining the growth of density
fluctuations, $\delta$, in terms of $\Omega$s. 
We can expand the $\dot{H}+H^2$-term in terms of $\delta$ and then 
the whole RHS of Eq. (\ref{gendenspe}) as
\be{genrhsexp}
3{1+\delta\over\bH^2}\Big((\dot{H}+H^2)-(\dot{\bH}+\bH^2)\Big)
\equiv 3(1+\delta)\sum_{n=1} c_n\delta^n.
\ee
Note that there is no constant term in the expansion.

Expanding the perturbation according to Eq. (\ref{deltaexp}) the linear
equation is then
\be{genlinear}
D_1''+(2+{\dot{\bH}\over \bH^2})D_1'+3 c_1D_1=0,
\ee
the second order equation
\be{gen2nd}
D_2''+(2+{\dot{\bH}\over \bH^2})D_2'-\frac 83 (D_1')^2+3c_1D_2
+6(c_1+c_2)D_1^2=0
\ee
and further orders are easily found. For example, in the standard case
we see that 
\bea{genestandc}
c_1 & = & -\frac 12 {\kappa^2\brhom\over\bH^2}=-\frac 12 {\Omega_M\over\Omega_M+a^3\Omega_{\Lambda}}\\
c_i & = & 0,\ i=2,3,....
\eea
Using these, the linear and second order equations, Eqs (\ref{lin})
and (\ref{2ndord}) are easily reproduced.

Note that these expressions make only sense on large scales where the
non-standard evolution of the universe can be seen. On small scales
the evolution obviously must be according to the standard Einstein's equations.

\section{DGP}
As we have seen, 
the two equations that determine how density perturbations grow are
the Friedmann equation and the continuity equation. In the DGP-
scenario these are:
\bea{dgpeqs}
0 & = & \dot{\rho}+3H(\rho+p)\\
H^2 & = & H_0^2\Big({\Omega_M\over
a^3}+\Omega_R(2+\sqrt{1+{\Omega_M\over\Omega_R a^3}})\Big).
\eea

In the DGP-case we cannot use Eq. (\ref{rayrhs}) since 
the Einstein's equations on the brane are modified from the 
standard four-dimensional equations as is apparent from the
non-standard Friedmann equation. Another way of seeing the same thing 
is to consider in the DGP scenario the quantity appearing on the 
RHS of the Raychaudhuri's equation, Eq. (\ref{ray2}),
\be{dgphs}
\dot{H}+H^2=-\frac 32\kappa^2(1+{1\over
R\sqrt{\kappa^2\rho+{1\over R^2}}})
(\rho+p)+\kappa^2\rho+{2\over R}\sqrt{\kappa^2\rho+{1\over R^2}}+{2\over R^2}.
\ee
This is different from the result that one gets in the
standard case, $-\frac 32\kappa^2(\rho+3p)$, with the standard
Einstein's equations. In the DGP-scenario the Raychaudhuri's 
equation hence needs to be calculated directly using the general 
approach described in the previous subsection.

In the matter dominated DPG-scenario:
\be{dgphdot}
{1\over H^2}(\dot{H}+H^2)=
\frac{\Omega_M\,\left( {\sqrt{\Omega_R}} - {\sqrt{\frac{\Omega_M}{a^3} + \Omega_R}} \right)
        + 4\,a^3\,\Omega_R\,\left( {\sqrt{\Omega_R}} + 
       {\sqrt{\frac{\Omega_M}{a^3} + \Omega_R}} \right) }{2\,a^3\,
    {\sqrt{\frac{\Omega_M}{a^3} + \Omega_R}}\,
    {\left( {\sqrt{\Omega_R}} + {\sqrt{\frac{\Omega_M}{a^3} + \Omega_R}} \right) }^2
    }.
\ee
The coefficients $c_i$ are easily calculated and the 
first two are, again expressed in cosmological quantities:

\bea{dgpv2cs}
c_1 & = & -\frac{2\,\left( \Omega_M + \Omega_R\,{a}^3 \right) \,
     \left( \Omega_M + 4\,\Omega_R\,{a}^3 \right)  + 
    {\sqrt{\Omega_R}}\,{\sqrt{\Omega_R + \frac{\Omega_M}{{a}^3}}}\,{a}^3\,
     \left( 5\,\Omega_M + 8\,\Omega_R\,{a}^3 \right) }
{4{\left( \Omega_M + \Omega_R\,{a}^3 \right) }^2}\nonumber\\
c_2 & = & \frac{\Omega_M^2\,{\sqrt{\Omega_R}}\,{\sqrt{\Omega_R + \frac{\Omega_M}{{a}^3}}}\,
    \left(8\Omega_R\, a^3 -\Omega_M \right) }{16\,
    {\left( {\sqrt{\Omega_R}} + {\sqrt{\Omega_R + \frac{\Omega_M}{{a}^3}}}
        \right) }^2\,{\left( \Omega_M + \Omega_R\,{a}^3 \right) }^3}\nonumber.
\eea
Finally, we need to calculate the term appearing in front of the
$\delta'$-term in the perturbation equation (\ref{gendenspe}):
\be{dgpA}
2+{\dot{H}\over H^2}=\frac{1}{2} + \frac{3\,{\sqrt{\Omega_R}}}
   {2\,{\sqrt{\Omega_R + \frac{\Omega_M}{{a}^3}}}}.
\ee
Clearly, in the limit $\Omega_R=0$, all these expressions
reduce to the corresponding expressions in the Einstein-deSitter case.

\begin{figure}
\includegraphics[width=70mm,angle=-90]{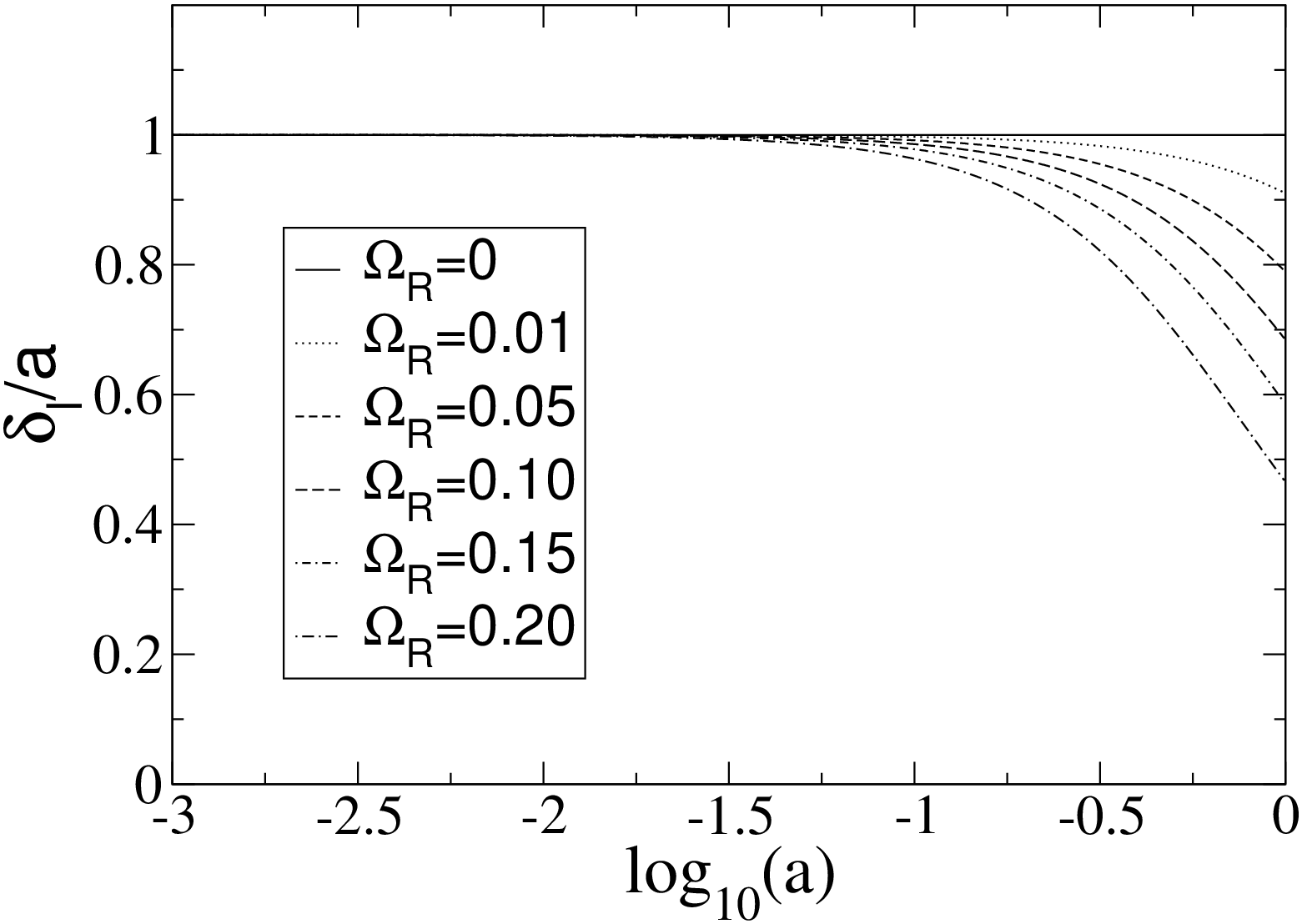}
\includegraphics[width=70mm,angle=-90]{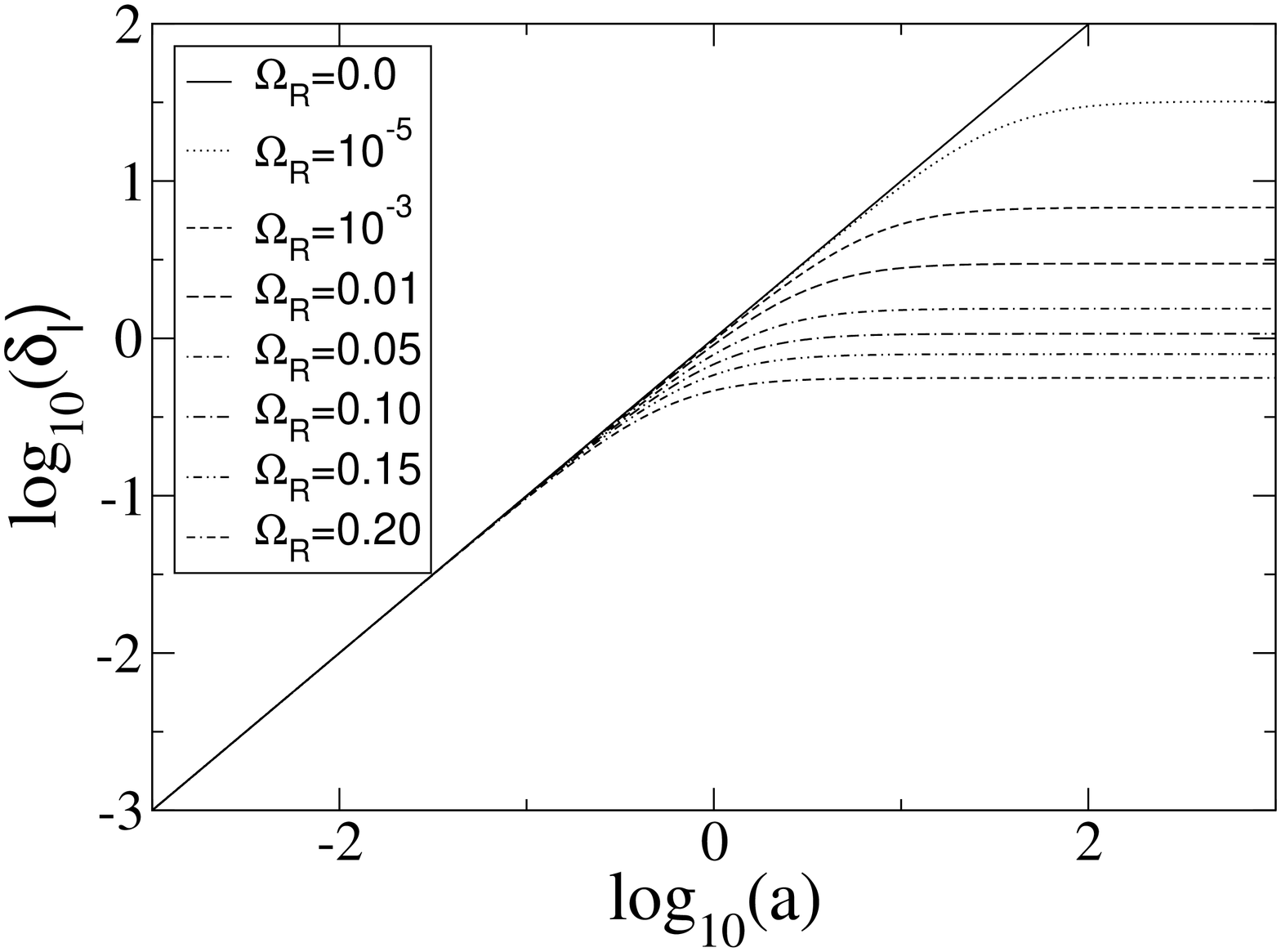}
\caption{Linear growth for different values of $\Omega_R$}
\label{dgpfig}
\end{figure}

\begin{figure}
\includegraphics[width=70mm,angle=-90]{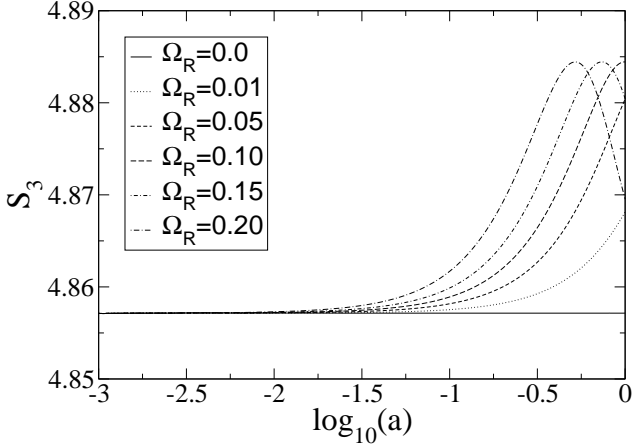}
\caption{Non-linear growth for different values of $\Omega_R$}
\label{dgpfig2}
\end{figure}

We can now study the growth of perturbations 
numerically. The initial
conditions are chosen such that at $a=10^{-3}$, the
standard exponential solution, $D_1\sim\exp(\eta)$, is reached. 
In calculating the second order perturbation, initial conditions are
chosen such that the standard solution, constant $S_3=\frac{34}{7}\approx
4.86$ is valid from the beginning.
In Fig. \ref{dgpfig}
the linear growth factor (and the linear growth factor normalized with 
the scale factor) for different values of $\Omega_R$, 
with $\Omega_M$ then determined by Eq. (\ref{flatnorm}), are shown 
as a function of the scale factor. The non-linear growth is shown in Fig.
\ref{dgpfig2}.

The general form of the linear growth factor is similar to the
cosmological constant case but here the effect is even more
pronounced. For the preferred value $\Omega_R=0.17$ \cite{dgp},
the growth of linear fluctuations up to now is suppressed by a factor
of $0.54$ compared to the EdS-case. In the $\Lambda$-universe
with $\Omega_{\Lambda}=0.7$, the suppression is $0.78$. Hence,
there will be significantly less structure growth on large scales
in the DGP-scenario than in a $\Lambda$-cosmology. Furthermore,
the suppression begins earlier in the DGP-scenario as is visible
from Fig. \ref{dgpfig2}.

The second order perturbation, or $S_3$, also grows differently
from the $\Lambda$-cosmology. The value of $S_3$ starts to grow
at earlier times and varies more than in the standard
$\Lambda$-universe.
However, the variation is still observationally insignificant being
at best less than one percent.

\section{Cardassian models}
In this section will consider the growth of gravitational
instabilities in the Modified Polytropic Cardassian model described by Eq.
(\ref{mpcequ}). The relevant equations in a matter dominated universe are
\bea{MPCfried}
0 & = & \dot{\rho}+3H\rho\\
H^2 & = & \kappa^2\,\rho{\Big( 1 + ( 1 + {z_{eq}})^{3(1-n)q}
          (\frac{\rho}{{\rho_0}})^{(n-1)q}\Big) }^{\frac{1}{q}}
\eea
From these all the relevant quantities can be calculated straightforwardly.

The $\dot{H}+H^2$-term is 
\be{mpcdoth}
{1\over H^2}(\dot{H}+H^2)=
-\frac 12 
\frac{1+ 
      (1+z_{eq})^{3(1-n)q}(3n-2)(\frac{\rho}{\rho_0})^{(n-1 )q}}
{1+(1+z_{eq})^{3(1-n)q}(\frac{\rho}{\rho_0})^{(n-1)q}}
\ee
and the first two $c_i$ coefficients
\bea{mpccs}
c_1 & = & -\frac 12 \frac{1+(4n-2+3(n-1)^2q)X+n(3n-2)X^2}
{\Big(1+X\Big)^2}\\
c_2 & = & \frac X4 (1+X)^{-3}
(n-1)\Big[\Big((1-n)q-1\Big)\Big(1+3(n-1)q\Big)\\
& & + \Big(2-4n-(n-1)(9n-5)q+3(n-1)^2q^2\Big)X+n(2-3n)X^2\Big],
\eea
where we have defined $X\equiv ((1+z_{eq})a)^{3(1-n)q}$
in order to shorten the otherwise lengthy expressions.

In the original Cardassian case with q=1, these expressions simplify to

\bea{carcoeffs}
c_1 & = & {-\frac 12+n(1-\frac 32
n)(1+z_{eq})^{3(1-n)}a^{3(1-n)}\over
1+(1+z_{eq})^{3(1-n)}a^{3(1-n)}}\\
c_2 & = & {\frac 12 n(n-1)(1-\frac 32
n)(1+z_{eq})^{3(1-n)}a^{3(1-n)}\over 1+(1+z_{eq})^{3(1-n)}a^{3(1-n)}}.
\eea
Again, as a check it is easy to see that with $n=0$, we recover
the standard coefficients (\ref{genestandc}).

Finally, in order to study the growth of perturbations, we need the
coefficient of the $\delta'$ term:
\be{mpcAterm}
2+{\dot{\bH}\over H^2}=2-\frac 32{1+n(1+z_{eq})^{3q(1-n)}a^{3q(1-n)}\over
1+(1+z_{eq})^{3q(1-n)}a^{3q(1-n)}}.
\ee

We can now calculate the growth of perturbations in the original
Cardassian scenario. Initial conditions are chosen like in the DGP
scenario.

The solution to the linear equation in the original Cardassian model
can be expressed in terms of the hypergeometric function. The growing 
part of the solution is found to be
\be{card0sol}
D_1(x)=e^x \hyper \Big[1,\frac{2+3n}{6(1-n)},{11-6n\over
6(1-n)},-e^{3(1-n)x}\Big],\ n<\frac 23.
\ee
For the general case, no such solution is found.

\begin{figure}
\includegraphics[width=70mm,angle=-90]{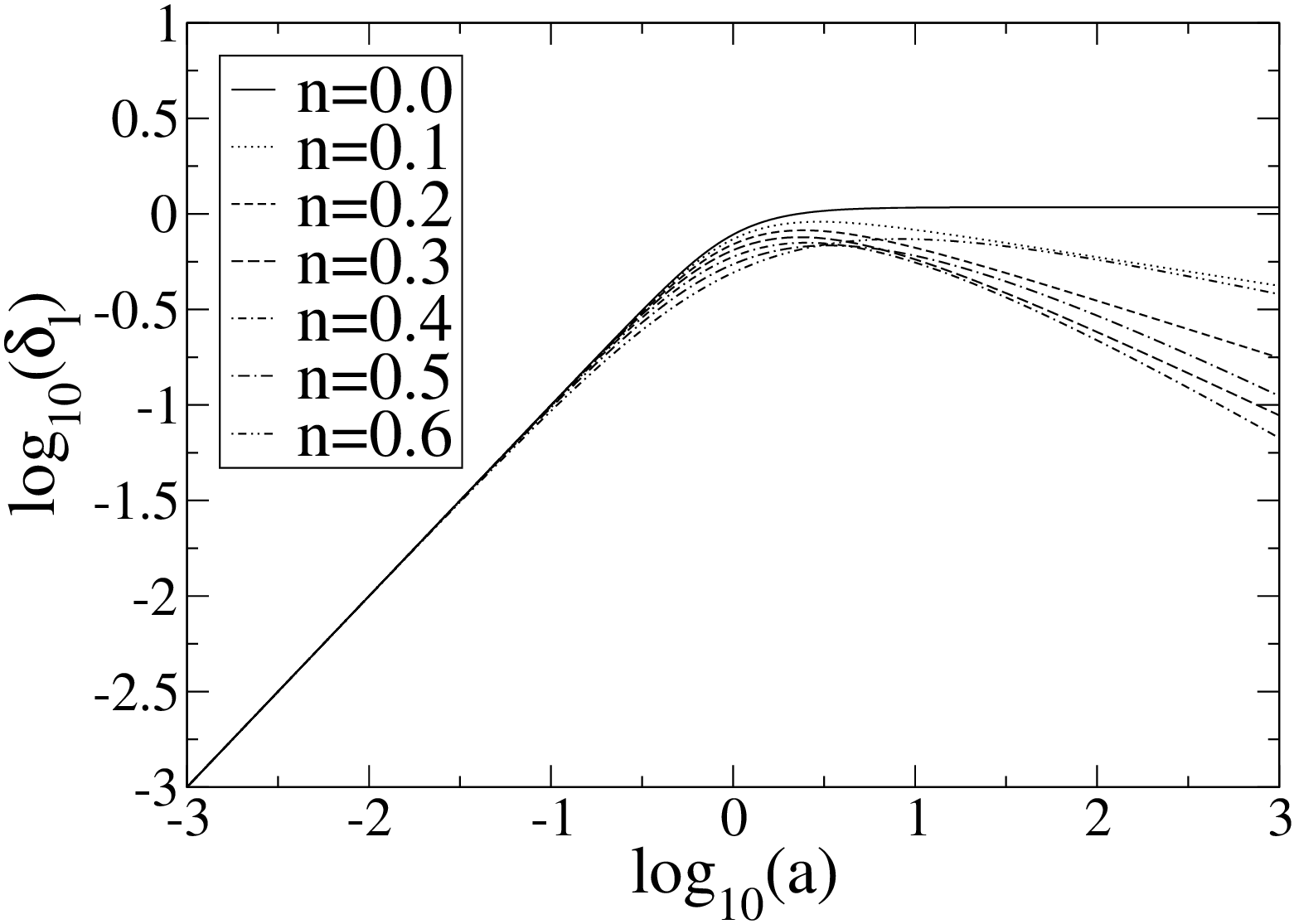}
\includegraphics[width=70mm,angle=-90]{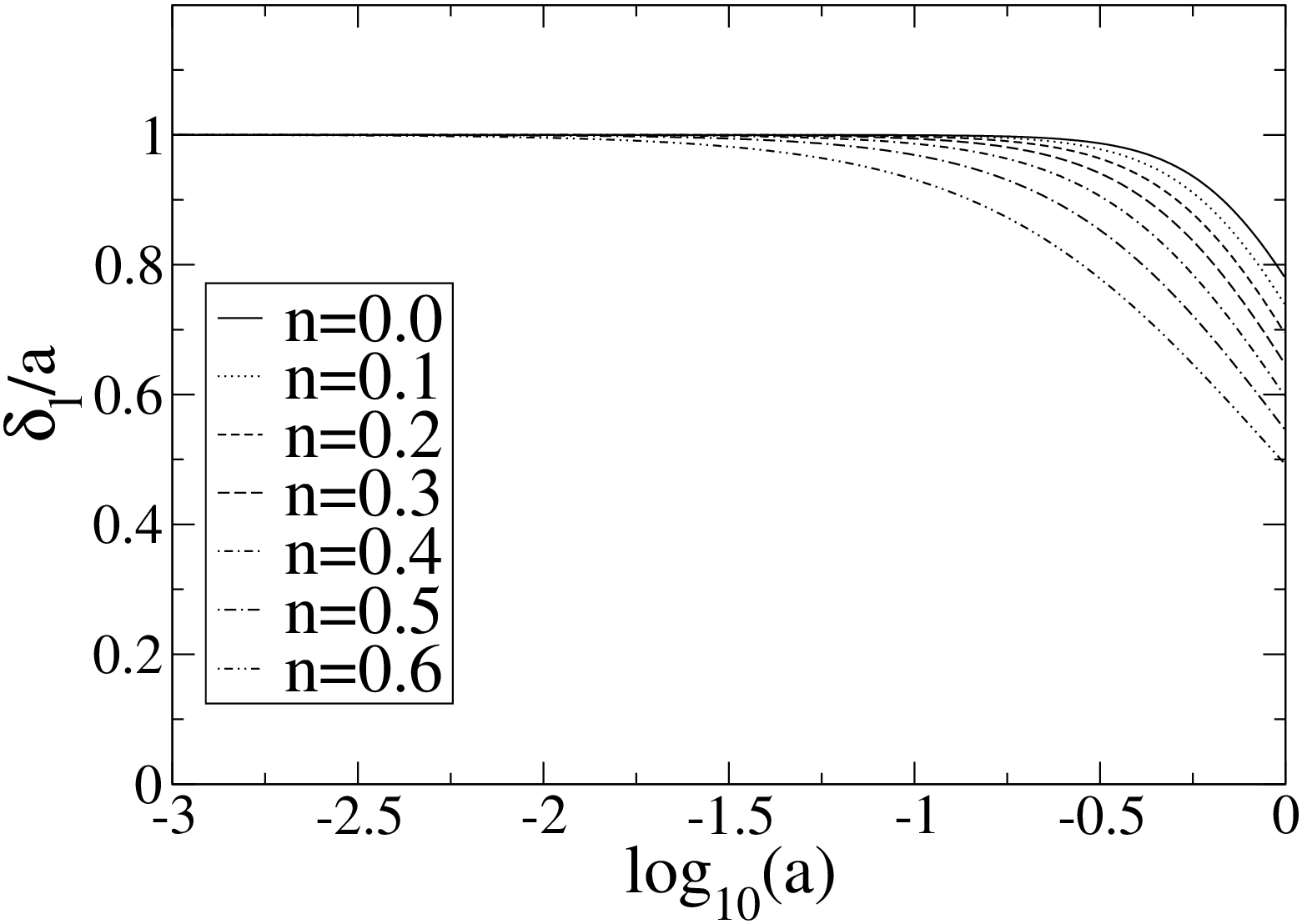}
\caption{Linear growth for different values of $n$, $q=1$}
\label{mpcfig1}
\end{figure}   

\begin{figure}
\includegraphics[width=70mm,angle=-90]{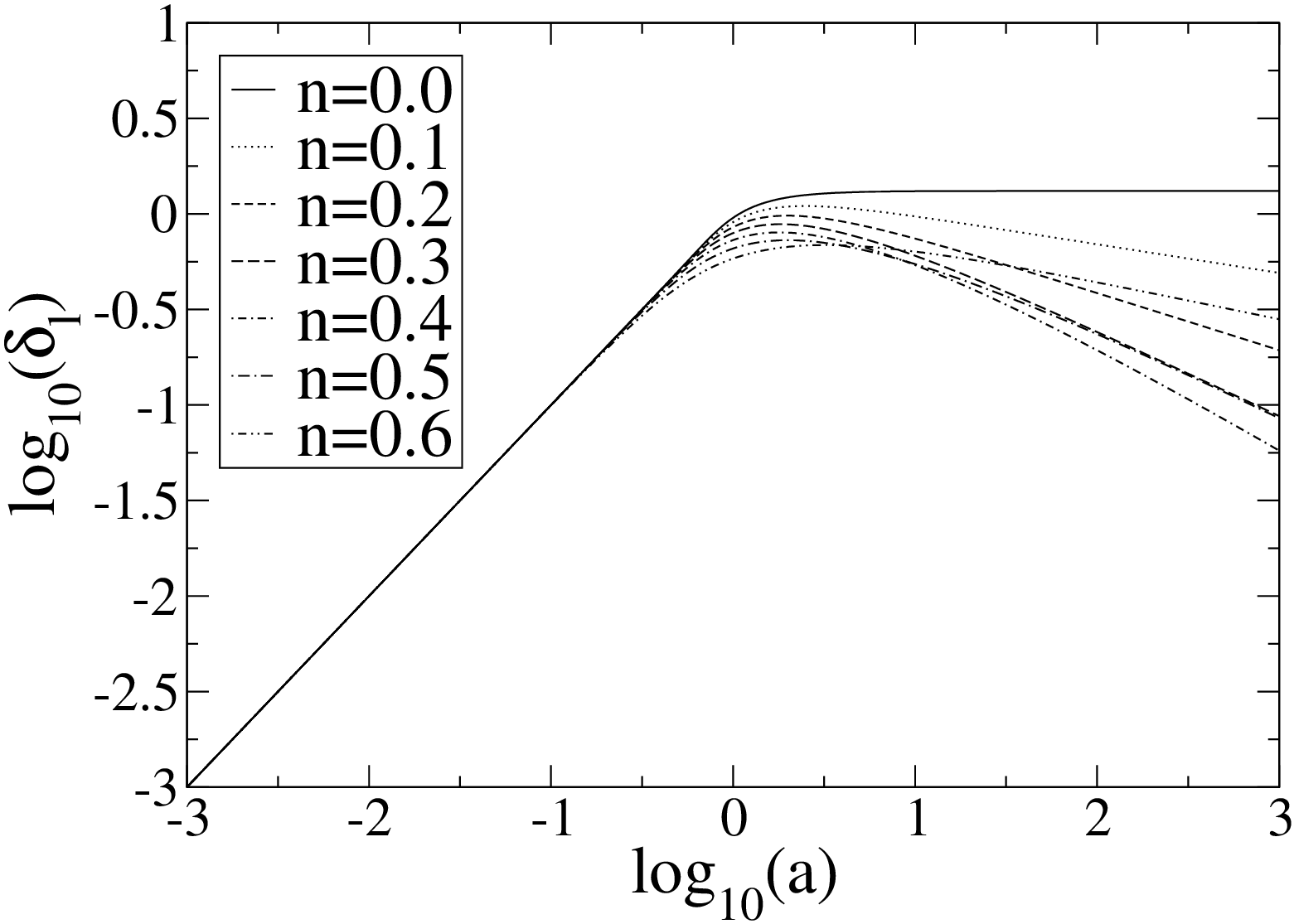}
\includegraphics[width=70mm,angle=-90]{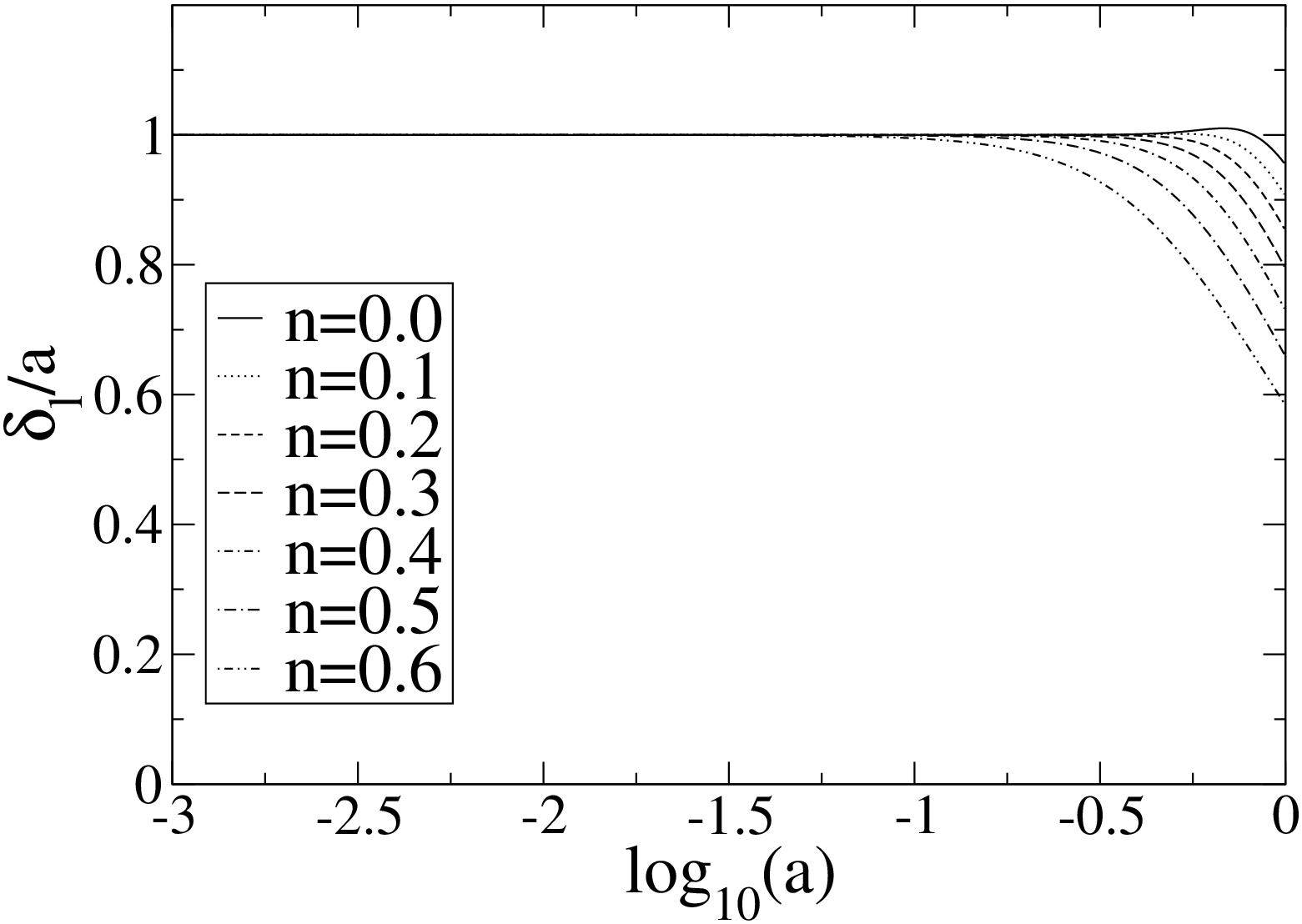}
\caption{Linear growth for different values of $n$, $q=2$}
\label{mpcfig2}
\end{figure}   

\begin{figure}
\includegraphics[width=70mm,angle=-90]{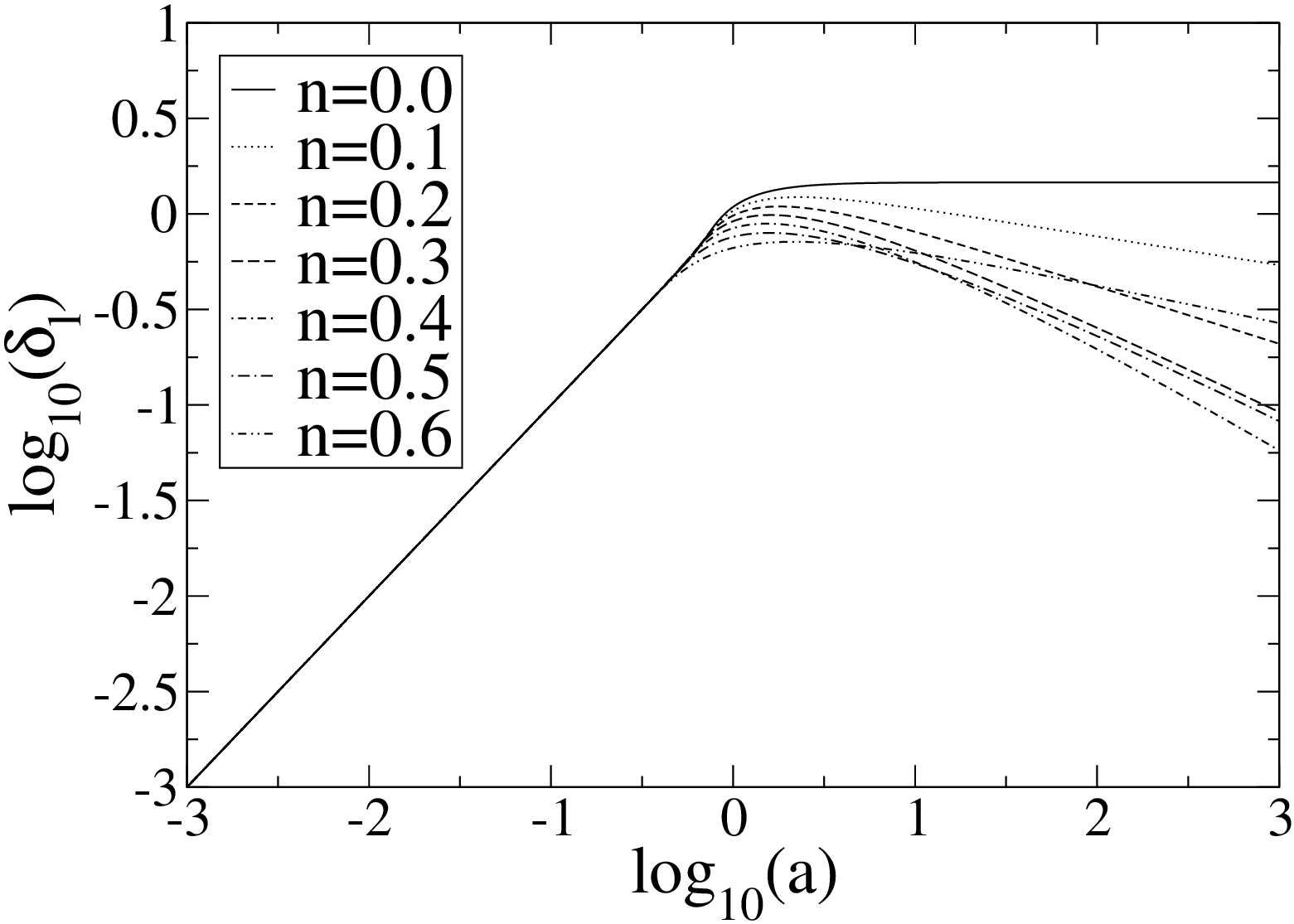}
\includegraphics[width=70mm,angle=-90]{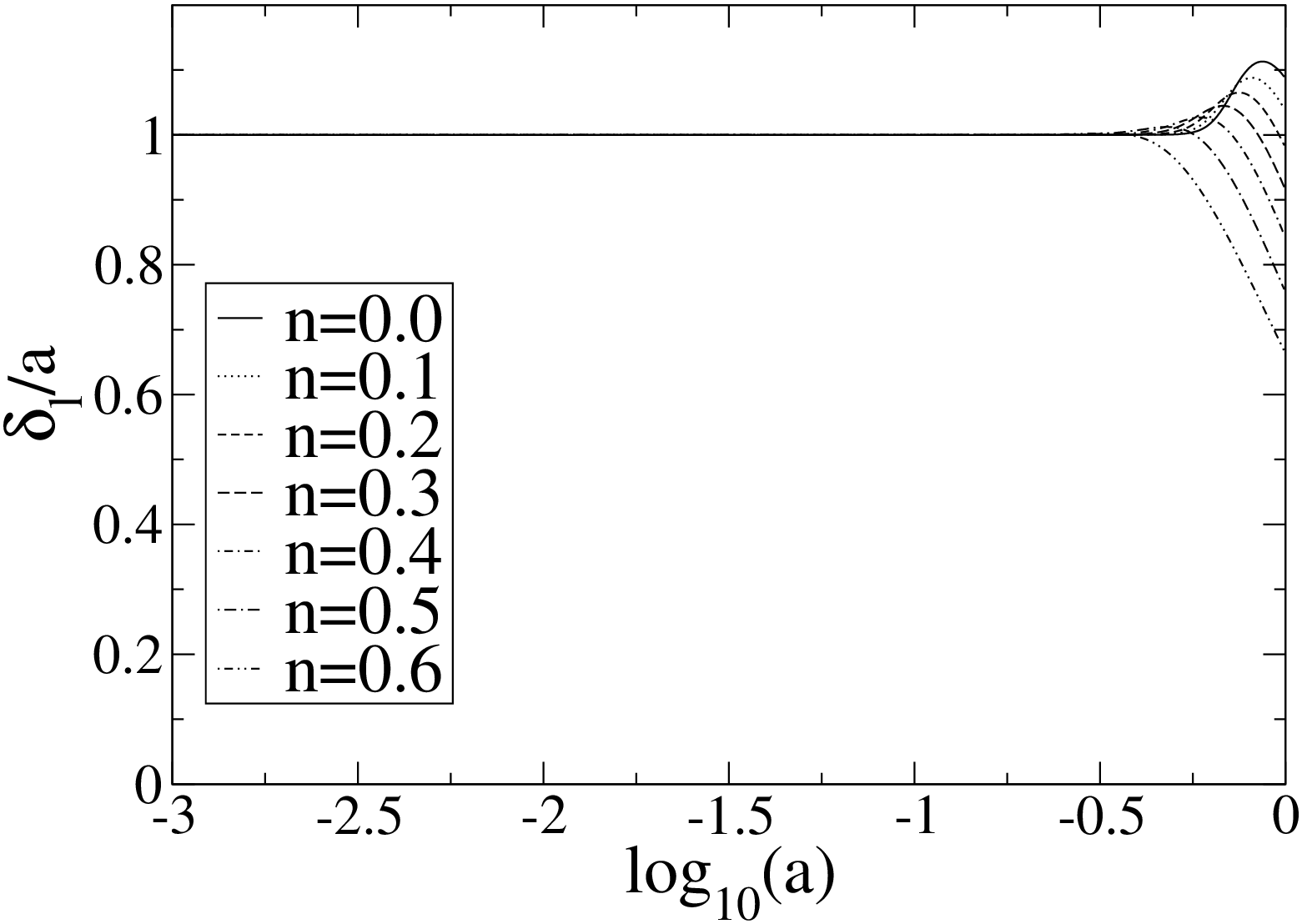}
\caption{Linear growth for different values of $n$, $q=5$}
\label{mpcfig3}
\end{figure}   

\begin{figure}
\includegraphics[width=70mm,angle=-90]{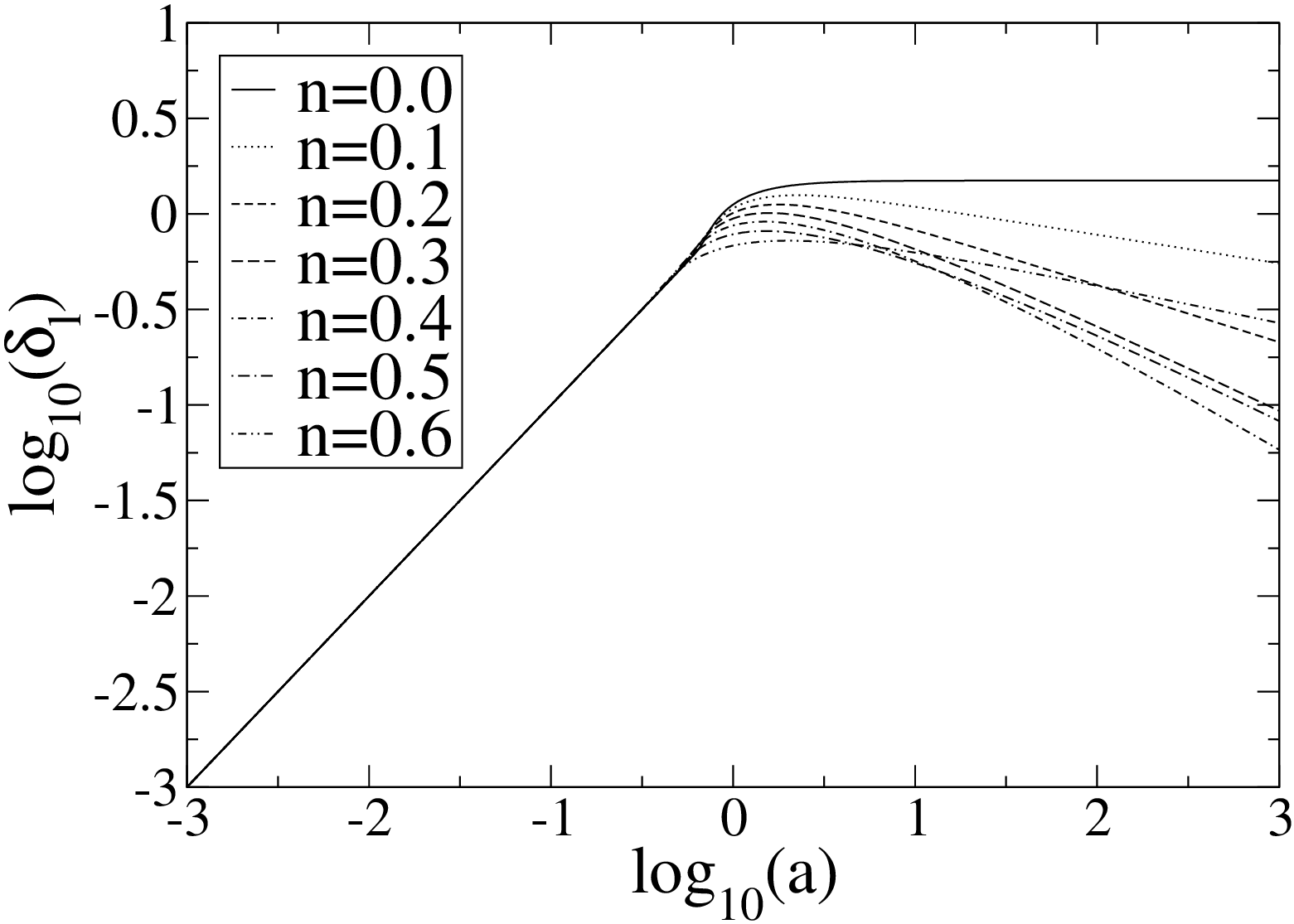}
\includegraphics[width=70mm,angle=-90]{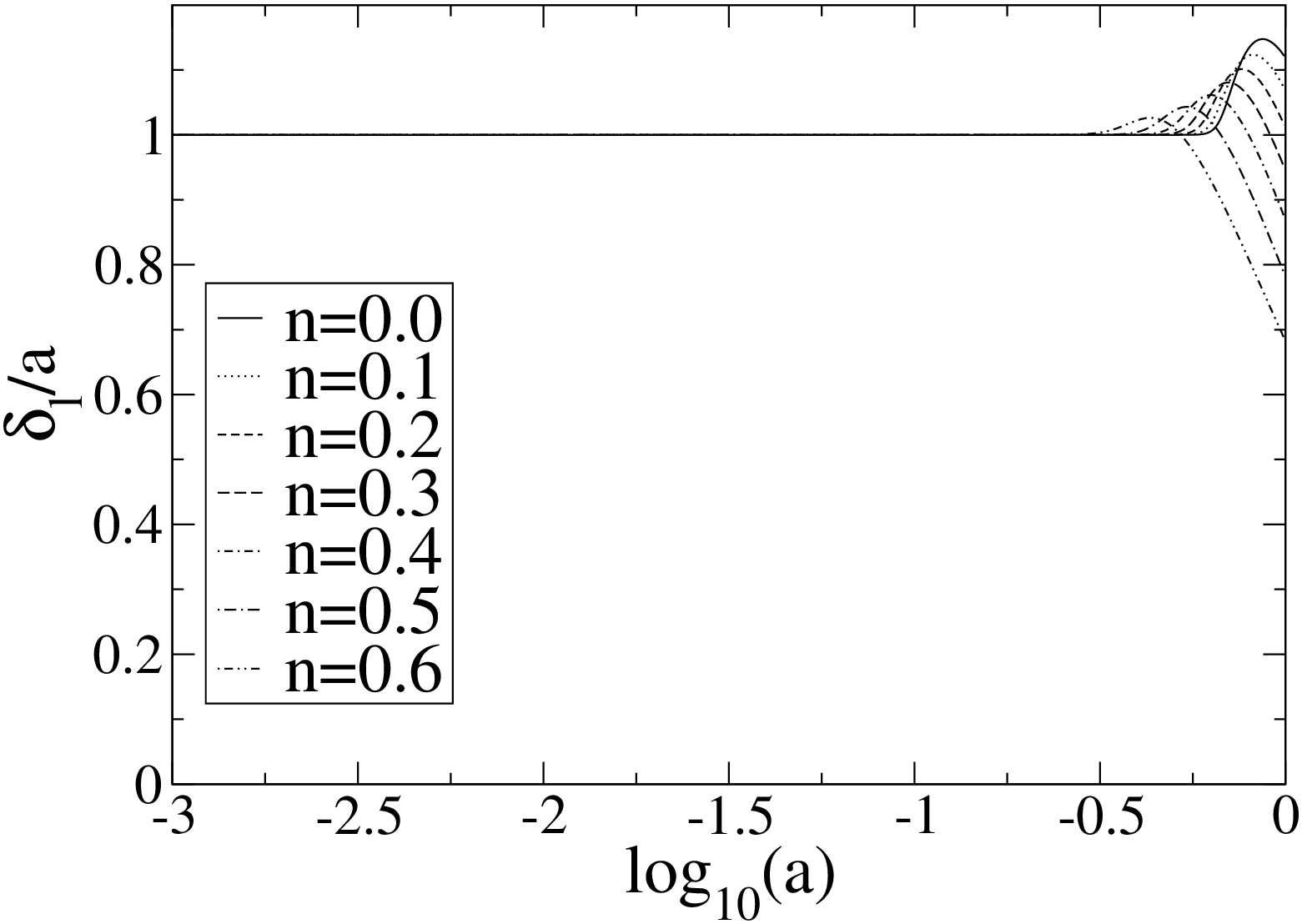}
\caption{Linear growth for different values of $n$, $q=10$.}
\label{mpcfig4}
\end{figure}

\begin{figure}
\includegraphics[width=70mm,angle=-90]{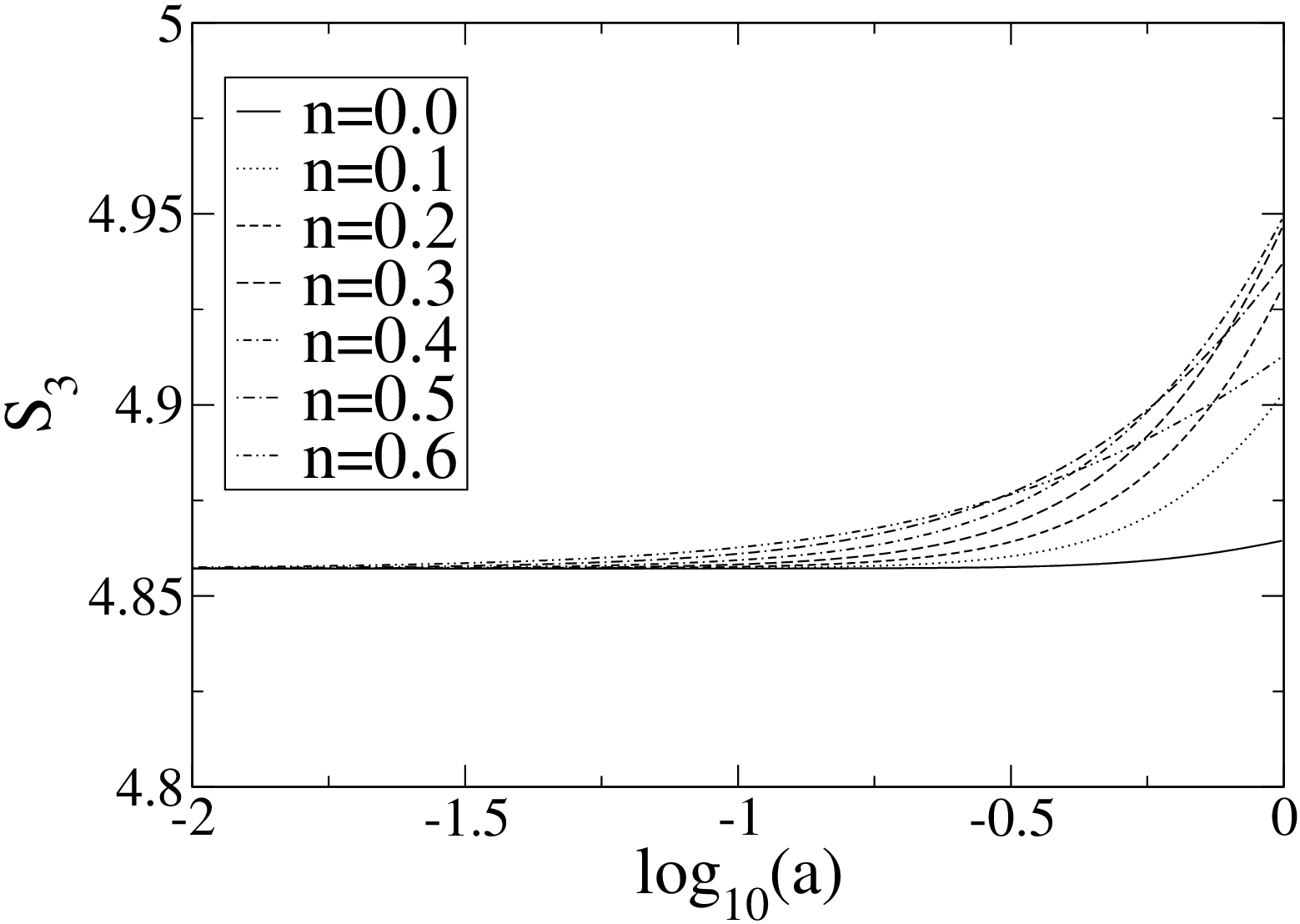}
\includegraphics[width=70mm,angle=-90]{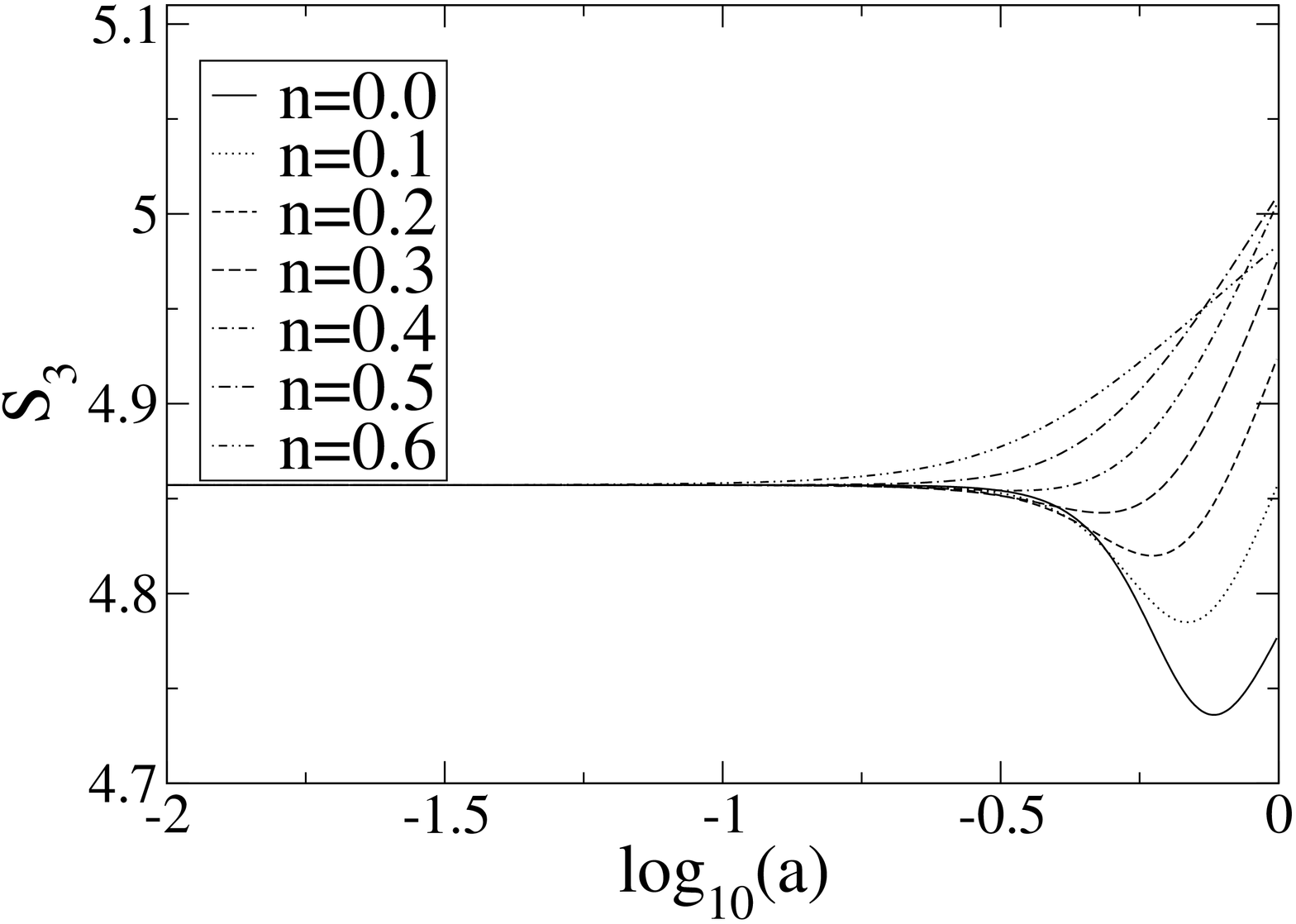}
\caption{Non-linear growth for different values of $n$, $q=1,\ 2$.}
\label{mpcfig5}
\end{figure}   

\begin{figure}
\includegraphics[width=70mm,angle=-90]{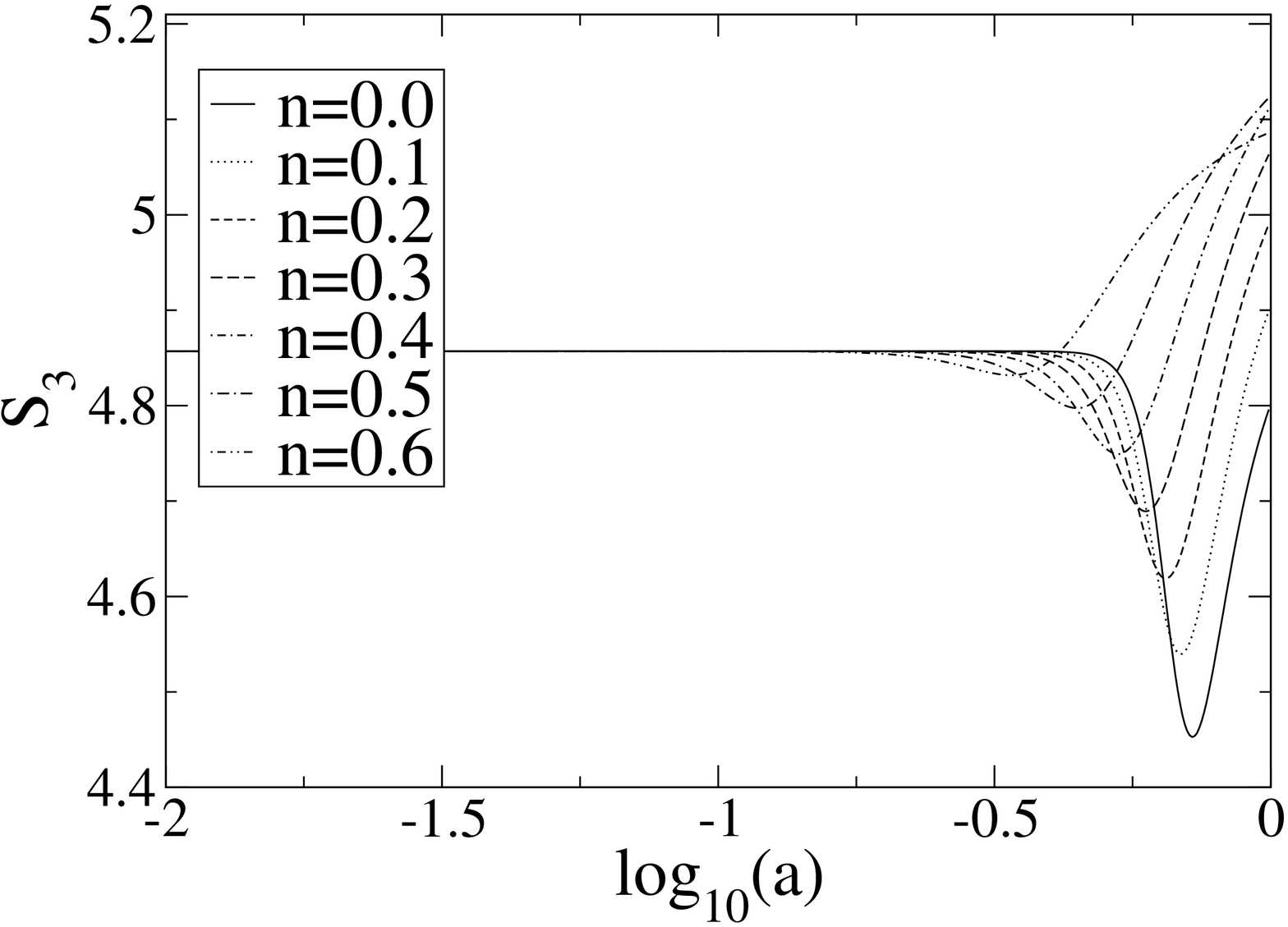}
\includegraphics[width=70mm,angle=-90]{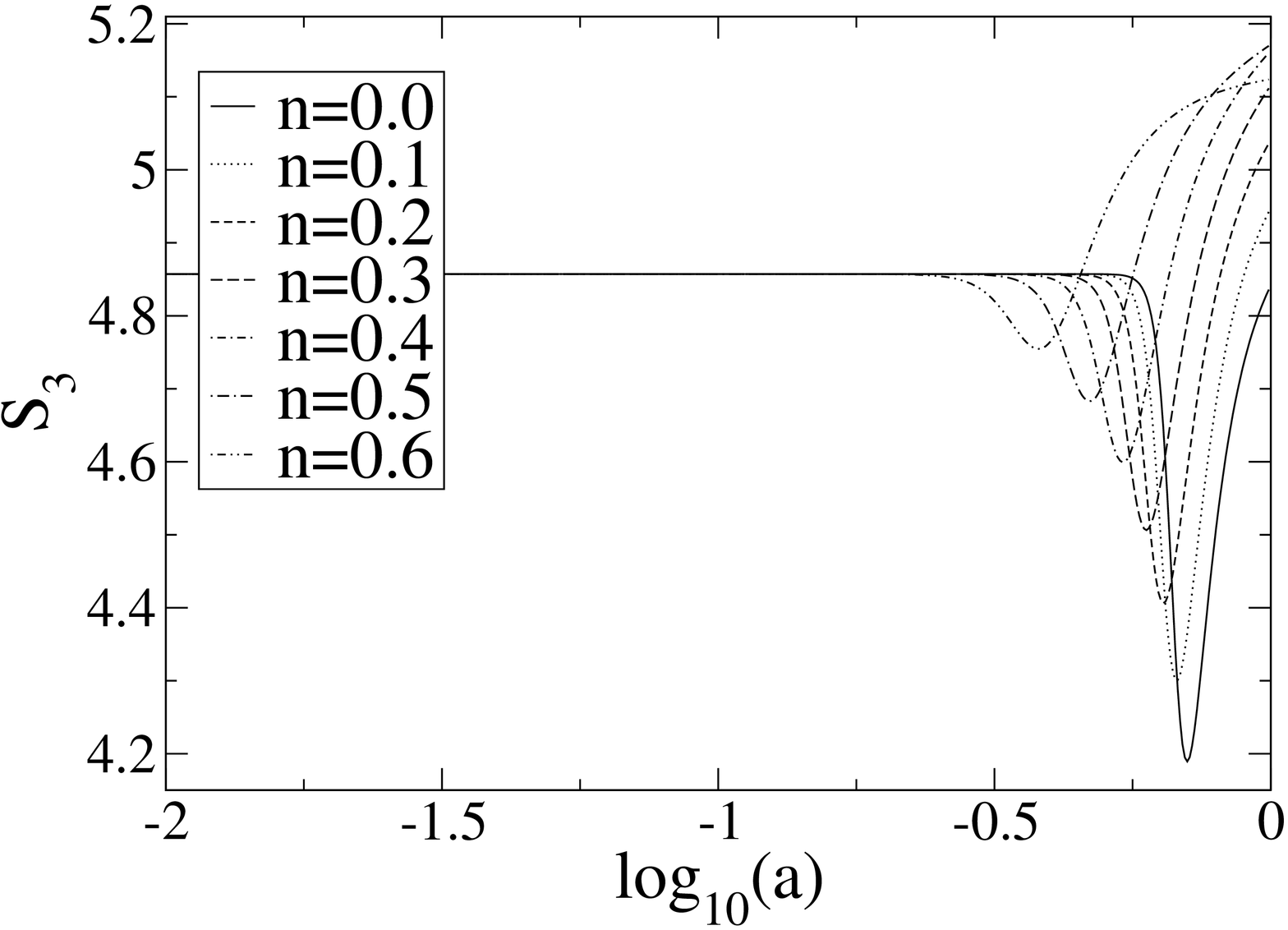}
\caption{Non-linear growth for different values of $n$, $q=5,\ 10$.}
\label{mpcfig6}
\end{figure}

We have plotted the linear and non-linear growth factors for
$n=0,0.1,...,0.6$ and $q=1,2,5,10$ as a function of $a$ in 
Figs \ref{mpcfig1}-\ref{mpcfig6}.
The linear growth factor is presented both on a logarithmic scale and
scaled by the scale factor. In plotting these figures, we have
assumed that $\Omega_M^{obs}=0.3$, which then sets the 
value of $z_{eq}$ according to Eq. (\ref{mpcobs}).

From the figures it is clear how the Cardassian scenario is
fundamentally different from the DGP-model and a $\Lambda$-cosmology.
Looking at the behavior of $\delta_l$
with $n>0$ (since $n=0$ corresponds to a cosmological constant)
the linear growth factor actually begins to decrease in the future \ie
areas of higher density begin to rarify. This is true
regardless of the value of $q$. Near the turning point
there is some interesting behavior as is visible from
the plots of $\delta_l/a$. With larger values of $q$ we see that
before the growth factor starts to decrease, structures grow
more rapidly compared to the standard case. This was not
seen in neither of the other models studied in this paper.
The value of $\delta_l/a$ at present is dependent on the value of
the parameters but can again be as small as $0.5$.

In Fig. \ref{mpcfig1} $q=1$ and we are hence looking at the 
original Cardassian model. This is of interest since in this
case we can understand some of the features by analytical 
considerations. For example, we see that the value of $n$
that corresponds to the slowest growth is $n\approx 0.4$.
If one considers the equation determining linear growth in more
detail, it can be seen that in the limit $a\rar\infty$,
the slowest growth actually corresponds to $n=\frac 49$ with $D_1(a)\sim 
a^{-\frac 23}$ as is shown in the Appendix.

Looking at the figures depicting the change in $S_3$, Figs
\ref{mpcfig5} and \ref{mpcfig6},  we see that
in the Cardassian model the evolution of the value of $S_3$
is strongly dependent on the value of $q$. The shape
and the magnitude of variation changes with $q$ so that
with larger $q$ one sees larger variations. 

In the original Cardassian scenario the values of $S_3$ grow
compared to the EdS scenario.
The overall shape of the curves is similar to that of the standard
$\Lambda\neq 0$ universe described in Sec. \ref{lamuni}. However,
here the scale of the change in $S_3$ is much larger than in
the standard $\Lambda\neq 0$ universe. The growth of non-linearities 
seems to be fastest for $n\approx 0.3$. Analytical considerations, 
see Appendix, show that in the large $a$ limit, the value of $n$ 
corresponding to fastest growth at large $a$ is again $n=\frac 49$. 
The magnitude of change in $S_3$ is approximately two percent with
$n=0.3$, which, although larger than in the DGP-scenario or
$\Lambda$-cosmology, is out of reach of present day experiments.

However, as $q$ is increased, also the variation of $S_3$ grows.
For example, with $q=10$, we see that $S_3$ can vary more than
$10$ percent from its EdS value. Also with $q\gsim 2$, the
values of $S_3$ undergo a change from values smaller than $S_3^{EdS}$
to values larger than $S_3^{EdS}$.


\section{Observational constraints}

As mentioned in the introduction, the above non-standard models
have been tested mostly against observations of SNIa and CMB.
The analysis presented
in this paper allows for a further comparison with constrains coming from 
linear and non-linear aspects of structure formation, 
\ie{\hskip-5pt}: $\sigma_8$ normalization, cluster number density 
and skewness of the density field. The shape of the initial spectrum
 of fluctuations $P(k)$ as a function of wavenumber $k$, 
could also depend strongly on cosmological parameters
(see eg Efstathiou, Sutherland \& Maddox 1990). But this requieres
additional ingredients to our models: ie 
 knowledge of the primordial spectrum and the matter content. 
Here we concentrate only in aspects that relate to
the matter dominated evolution, assuming some given
 initial shape $P(k)$.
We briefly sketch how this comparison can be done 
and leave a more direct parameter estimation for future
work.

\subsection{Amplitude normalization: $\sigma_8$}

\begin{figure}
\includegraphics[width=100mm]{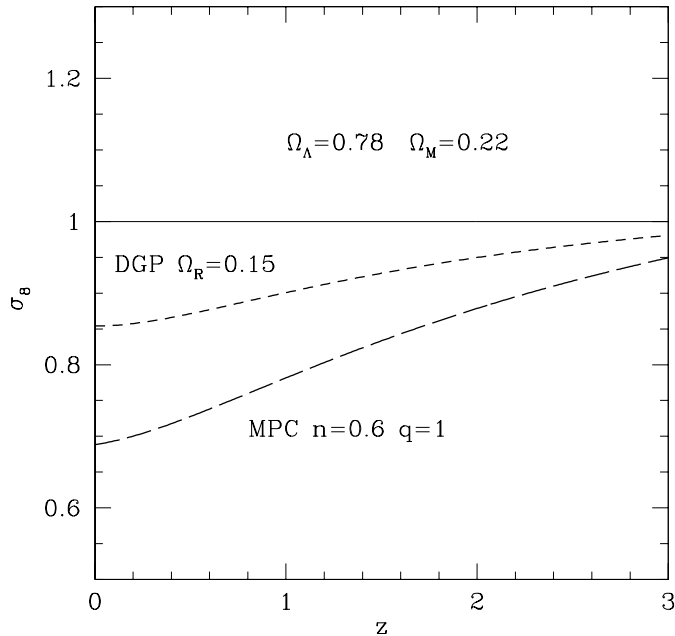}
\caption{Value of $\sigma_8$ in non-standard cosmologies,
with $\Omega_M=0.22$, 
relative to the value for the $\Omega_\Lambda=0.78$, $\Omega_M=0.22$
standard cosmology with $\sigma_8=1$.}
\label{plotsig8}
\end{figure}

The amplitude of density fluctuations is commonly characterized
by the linear value of the rms fluctuations on a sphere of
$8 \Mpc$. Observationally its value seems to be of order
unity at $z=0$. Form the analysis in the previous sections
it is straight forward to predict how different $\sigma_8$ 
should be for a given cosmological model. One way to do this
is to fix the normalization of all models to be equal at high
red-shifts and see how different they are at $z=0.5$.\footnote{
In principle, this is similar to a CMB normalization,
but note that in general a CMB normalization also requires
some additional information on the shape of the spectrum, which 
might vary from model to model. Thus, here we choose to 
illustrate how $\sigma_8$ changes due to the linear
growth from a given fixed shape of the spectra.}
Figure \ref{plotsig8} illustrates this point. We show the value of 
$\sigma_8$ normalized to  the $\Lambda$ model as a function
of red-shift for DGP and MPC ($n=0.6$ and $q=1$) cosmologies, 
all with same value of $\Omega_m=0.22$. 

Values of $\sigma_8$ for other parameters can readily be obtained
by comparing Figure \ref{stfig} with Figure \ref{dgpfig} and
Figures \ref{mpcfig1}-\ref{mpcfig4}. Note how for MPC with $q>1$ we
can also get $\sigma_8>1$.

Current observational constraints of $\sigma_8$ usually come with several
different assumptions. In most cases a shape for the linear spectrum is
assumed to get an estimation, for example from cluster abundances or
normalization of CMB fluctuations. If the primordial spectrum of
fluctuations is not scale invariance or a simple power-law, such 
extrapolations could yield misleading values of $\sigma_8$ 
(e.g. see \cite{barriga} and references therein). 
Large scale structure in local galaxy catalogues also yield
values  $\sigma_8 \simeq 1$, but
one needs to quantify the bias $b$, or how
well the selected galaxies trace the underlying mass distribution (see e.g.
Gazta\~naga 1995). Recent CMB results from the {\it WMAP} (Spergel et al. 2003)
mission fit to the (power-law) $\Lambda$CDM
cosmology find: $\sigma_8=0.9 \pm 0.1$ ($\sigma_8=0.84 \pm 0.04$ with a running
index, \cite{spergel}). This value has been extrapolated
from $z\simeq 10^3$ to $z=0$ by using the linear growth factor $D_1(z)$ 
in Eq. (\ref{lamunisol}). Significantly smaller values 
($\sigma_8\simeq 0.6-0.7$) have been found at low red-shifts by weak lensing
(see e.g. \cite{jarvis}) and velocity fields (e.g. Willick \& Strauss 1998). 
This apparent discrepancy, if real, could be easily accounted for by using
the DGP or MPC cosmological models.

\subsection{The number counts of galaxies}

\begin{figure}
\includegraphics[width=100mm]{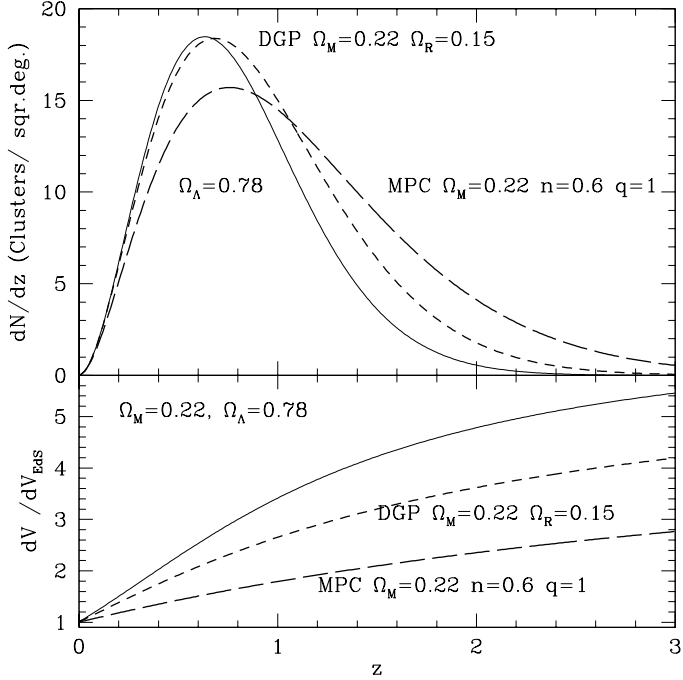}
\caption{Bottom: Comoving volume element as a function of red-shift
normalized to the EdS model. The short and long dashed line shows DGP 
and MPC predictions for 
$\Omega_M=0.22$ ($\Omega_R=0.15$) and $h=0.71$. The continuous
line corresponds to the $\Lambda$ cosmology with
$\Omega_M=0.22$ ($\Omega_\Lambda=0.78$) and $h=0.71$.
Top: Number density of predicted clusters per square degree 
with mass $M> 2\times 10^{14} M_{solar}$ for the same models. All
models are normalized to $\sigma_8=1$ at $z=0$.}
\label{PS}
\end{figure}     

Press \& Schechter formalism (Press \& Schechter 1974) 
and its extensions (see e.g. Bond et al. 1991; Lacey \& Cole 1993)
predict the evolution of the mass function of collapse
objects. These predictions are based on assuming Gaussian initial conditions
and the spherical collapse model (which is closely related to the
shear-free approximation used in Eq. (\ref{ray})). 
Despite the apparent limitations of these
assumptions, comparison with realistic simulations show good agreement 
(see a recent review by Cooray \& Sheth (2002)). 
In the standard Press-Schechter formalism,  the comoving number 
density of collapsed objects (halos or clusters) of mass $M$ is

\begin{equation}
n(M) dM = - \sqrt{2\over \pi} 
\, \left({\delta_c \over \sigma}\right) {d \ln \sigma  \over 
d \ln M} \, \exp{\left(-{\delta_c^2 \over 2 \sigma^2}\right)} \, 
{\bar\rho \, dM \over M^2}
\label{eq:PS}
\end{equation}
where $\sigma=\sigma(R,z)$ is the linear rms fluctuation at the scale
$R$ corresponding to the mass $M= 4/3\pi R^3 \bar\rho$, and $\bar\rho$ is
the mean background. The value of $\delta_c$ corresponds to the value of
the linear over-density at the time of collapse, \ie
when the non-linear over-density becomes very large
$\delta \rightarrow \infty$. This value can be found by solving
Eq.(\ref{standenspe}).  For the standard Einstein-de Sitter case we have
$\delta_c \simeq 1.686$. We find little difference in $\delta_c$
for the non-standard cosmologies, so for simplicity we will use 
the EdS value in all cases.

To translate the above predictions into observable quantities, such as the
number density of clusters per unit red-shift, above a certain mass,
we need to integrate over the comoving volume element $dV/dz$, which
is also a strong function of the cosmology. Bottom panel of Figure
\ref{PS} shows the comoving volume element (normalized to EdS case)
in the DGP (dashed line), MPC (long dashed line) and 
$\Lambda$ cosmology (continuous line).
As can be seen in the figure, the $\Lambda$ cosmology has about 4 times
more comoving volume by $z\simeq 1.5$ than the EdS case, while DGP
and MPC are only 3 and 2 times larger. Despite the smaller volumes, DGP 
and MPC predicts
2  and 4 times more clusters at $z\simeq 1.5$ 
than the $\Lambda$ cosmology because of the
stronger freeze in the linear growth factor, 
 which can be seem by comparing Figure \ref{stfig}
to Figure \ref{dgpfig}. 

Note that  there are four important factors in the PS predictions:
i) $\Omega_M$, which relates a given cluster mass $M$ to $R$ in $\sigma(R)$
value, ii) the shape of the power spectrum, which determines the
$\sigma(R)$ curve, iii) the volume element, and iv) the growth factor
$D_1(z)$ which gives the red-shift variation of $\sigma$. In our
analysis we fixed i) and ii) to the $\Lambda$ cosmology with 
$\Omega_M \simeq 0.22$. Thus all differences in Figure
\ref{PS} are due to differences in volume and growth factors, which
mark the distinction between standard and non-standard cosmologies.

\subsection{The skewness $S_3$}

We have already shown the predictions for the normalized skewness,
$S_3$ in Eq. (\ref{skew}), in Figures \ref{stfig2}, \ref{dgpfig2},
\ref{mpcfig5} and \ref{mpcfig6} (see also Eq. (\ref{apps3}) 
in the appendix). These are the unsmoothed values of the skewness
at a single point. In practice, to relate to observations, one
needs to take into account smoothing effects. For a top-hat 
window of a given  radius $R$, smoothing results in a simple linear
correction that is given by the slope $\gamma$ 
of the variance smoothed at a radius $R$ 
(Juskiewicz, Bouchet \& Colombi 1993). In the standard cosmology:
$S_3= 34/7 + \gamma$ (eg see \cite{bcgs} and references therein). 
The smoothing correction should be the same for non-standard cosmologies,
as long as we work within the spherical collapse model 
(ie the shear-free approximation  in Eq. (\ref{ray})), which 
gives the exact leading order contribution to $S_3$ for a top
hat window (see e.g. Fosalba \& Gazta\~naga 1998; Bernardeau et al. 2002).
Thus, the difference between the standard predictions and the predictions
from non-standard cosmologies should not be affected by these smoothing
effects.  Besides smoothing effects, $S_3$ is also affected by systematic
uncertainties of biasing (see \S7.1 in Bernardeau et al. 2002).

Current estimations for $S_3$ (see \S8 in \cite{bcgs}) agree with the
standard predictions but with large uncertainties, of order $20\%-30\%$.
Thus current observations on $S_3$ can not be used to separate the 
different models. Next generation catalogues, such as the Sloan Digital
Sky Survey, are expected to
reduce the sampling variance uncertainties to $5\%$ (see e.g.
Szapudi, Colombi \& Bernardeau 1999; Verde et al. 2000).
Thus, after removing biasing uncertainties, future measurements on $S_3$
could be used to constraint some of the MPC parameters 
(e.g. see Figs. \ref{mpcfig5} and  \ref{mpcfig6}).

\section{Conclusions}

In this paper we have studied the growth of large scale structure in
non-standard cosmological scenarios that can undergo
acceleration without having an explicit cosmological constant.
In order to do this, we have extended the standard spherical 
collapse formalism to account for the non-standard equations.
All that is needed is the Friedmann equation and the assumption
that the energy density of matter is conserved. The applicability
of our approach is limited to very large scales, where
Einstein's equations can be modified. Of the two studied scenarios, 
the DGP-model is more attractive
since it is based on physics coming from brane cosmology
whereas the MPC-model is a more phenomenological approach.
In a flat universe the DGP-model has only a single free parameter, 
$\Omega_M$, but this model is still found to be in accordance with 
the current observations on large scale structure. The MPC model
has two additional parameters so that it is less surprising that
observations on large scale structure do not rule out such a model.

In the DGP- and MPC- models it was found that linear growth
is inhibited due to the accelerated expansion, like in the
standard $\Lambda$-cosmology. Depending on the values of the
parameters, the suppression at present compared to the EdS-case
can be as much as 50\%. The non-linear growth was studied by 
considering the value
of the skewness. For DGP, the deviation from the EdS-value is larger
than in the $\Lambda$-case but still only of the order of one
percent. In the MPC-scenario, however, the story is quite different 
and the value of skewness can vary up to $10$ percent.

In order to evaluate the significance of the non-standard evolution
of large scale perturbations, one obviously needs to relate the
predictions of the linear and non-linear growth rate to observations.
As was discussed in Section 6, the linear normalization of large
scale density fields can vary greatly. The value of $\sigma_8$ can 
be much smaller in the MPC- and DGP- scenarios when compared to the
standard $\Lambda$-cosmology with the same normalization
at high redshift.  Intriguingly, there is already some
observational indication of a small value of $\sigma_8$ at low
red-shifts but more observational data is needed in order to 
determine whether this effect is real. 
We also find that in the MPC-scenario, the linear
growth can actually be faster than in the EdS universe for
a limited range of red-shift. This is an interesting property
since models that have late time acceleration typically lead to
less linear growth at all times. In any case, observations
of $\sigma_8$ at low red-shifts, combined with the precision CMB data
coming from the {\it WMAP} (and in the future {\it Planck}) mission, will be 
a powerful probe of non-standard cosmological evolution.

Another quantity that is strongly affected by the non-standard
linear evolution is the number counts of objects. As it was shown
in Section 6, using the Press-Schechter formalism gives
strong predictions for the number of clusters that differ significantly
from one expects in a $\Lambda$-cosmology. Again, studying
the number counts of clusters at different red-shifts gives another
way to constrain non-standard cosmological models such as the
DGP- and MPC-scenarios.

The higher order statistics can also be modified by the non-standard
effects. In this paper we have considered the normalized skewness, 
$S_3$, whose evolution can be straightforwardly extracted in each model.
The skewness gives another probe of possible atypical cosmological
evolution. Typically, the effect on the skewness of different
cosmological scenarios, \eg $\Lambda$-cosmology and 
quintessence \cite{quinte}, is very small and totally unobservable.
However, as we have seen here (see also Gazta\~naga \& Lobo 2001), 
the effect can be much larger and can possibly be detected in 
the near future experiments such as the
Sloan Digital Sky Survey.

The growth of large scale structure is tightly bound to the 
overall evolution of the universe and hence to the particular
cosmological model.  As such, along with the SNIa and CMB data, 
it is a powerful tool to probe the space of possible
cosmologies. Current large scale structure data
agrees well with the two proposals
that explain late time acceleration without a cosmological constant.
This agreement is not trivial in the case of DGP, as there are no
additional free parameters, once we fit to the SNIa data or the value
of $\Omega_M$.  We have shown here how upcoming surveys of large scale  
structure can be used to further constrain these non-standard models
and differentiate then from the now standard $\Lambda$ model.


\section*{Acknowledgments}
One of us (TM) is grateful to the Academy of Finland for financial 
support (grant no. 79447) during the completion of this work.
EG and MM acknowledge support from
and by grants from IEEC/CSIC and the spanish Ministerio de Ciencia y
Tecnologia, project AYA2002-00850 and EC FEDER funding.
MM acknowledges support from a PhD grant from Departament d'Universitats,
Recerca i Societat de la Informacio de la 
Generalitat de Catalunya.
We are grateful to the 
Centre Especial de Recerca en Astrofisica, Fisica de Particules i Cosmologia
(C.E.R.) de la Universitat de Barcelona and IEEC/CSIC 
for their support.



\newpage
\appendix
\section{Growth of linear and second order perturbations in the
Cardassian model}
In this appendix, we consider the growth of linear and second order
fluctuations in the original Cardassian model, $q=1$, by analytical means.

\subsection{Linear growth}
The equation determining linear growth, Eq. (\ref{genlinear}), in the
Cardassian model in a scaled form is
\be{applin}
{d^2D_1\over dx^2}+\Big(2-\frac 32 {1+ne^{3(1-n)x}\over 1+e^{3(1-n)x}}\Big)
{dD_1\over dx}+\frac 32 {2n(1-\frac 32 n)e^{3(1-n)x}-1\over
e^{3(1-n)x}+1}D_1=0,
\ee
where $x\equiv\ln(a/a_C)=\eta-\eta_C$. 

It is now interesting to look at the large $\abs{x}$ limits.
Since we are interested in the range of values where $0<n<\frac 23$,
it is clear  that at, depending on the sign of $x$, the
exponential terms in Eq. (\ref{applin}) will either
dominate or be negligible in the large $x$ limit.
When $x$ is large and negative, \ie when $a/a_C\ll 1$,
Eq. (\ref{applin}) takes the form
\be{applinminus}
{d^2D_1\over dx^2}+{1\over 2}{dD_1\over dx}-\frac 32D_1=0,
\ee
\ie the standard form, which has the usual solution
\be{applinsol1}
D_1^{-}(x)=A_1 e^x+A_2 e^{-\frac 32 x}.
\ee

In the large positive $x$-limit, \ie when $a/a_C\gg1$,
it is obvious that the exponential terms will dominate
and so that Eq. (\ref{applin}) can be written as
\be{applin2}
{d^2D_1\over dx^2}+(2-\frac 32 n){dD_1\over dx}+3 n(1-\frac 32 n)D_1=0.
\ee
The solution to this equation is easily found and reads as
\be{applinsol}
D_1^{+}(x)=B_1e^{-\frac 32 nx}+B_2e^{(3n-2)x},
\ee
where $B_i$ are constants. Looking at the solutions, we see that
with $n=1$, the two solutions agree as expected. If $n<\frac 23$,
the linear growth of perturbations will at some point stop and
perturbations start to shrink.

The slowest growth rate at large $a/a_C$ is easily deduced
from $D_1^{+}(x)$ and occurs when 
$-\frac 32n=3n-2$ \ie when $n=\frac 49$ and hence $D_1^{+}(x)\sim e^{-\frac 23x}$.

\subsection{Second order perturbations}
The equation determining the growth of second order perturbations
in the large $x$ limit, in the negative large $x$ limit we 
obviously reproduce the standard scenario again, is
 from Eq. (\ref{gen2nd}),
\be{app2nd}
{d^2D_2\over dx^2}+(2-\frac 32 n){dD_2\over dx}+3 n(1-\frac 32
n)D_2-\frac 83 \Big({dD_1\over dx})^2+3n(n+1)(1-\frac 32 n)D_1^2=0.
\ee
The part that will dominate the
linear solution $D_1$ depends on $n$, if $n<\frac 49$, $D_1\sim \exp(-\frac
32 nx)$, where as if $\frac 49<n<1$, $D_1\sim \exp((3n-2)x)$.
Let us first assume that $n<\frac 49$, in which case by substituting
the appropriate solution of $D_1$ into Eq. (\ref{app2nd}) and
solving the resulting equation, we get
\be{app2ndsol1}
D_2(x)=\frac 12 (2+n)e^{-3nx}+C_1 e^{-\frac 32 nx}+C_2 e^{(3n-2)x}.
\ee
Hence, since $n<\frac 49$, we see that in the large $x$ limit, $D_2\sim
e^{-\frac 32 n x}$. Therefore, we expect that in this limit,
\be{apps3}
S_3\sim e^{\frac 32 nx}.
\ee
If we instead look at the values of $\frac 49<n<1$, we see that 
$S_3\sim e^{(2-3n)x}$, and hence the value of $n$ that gives the
largest effect rate of change of $S_3$ is, again, at $n=\frac 49$
with $S_3\sim e^{\frac 23 x}$.


\label{lastpage}

\end{document}